\documentstyle[12pt,a4]{article}
\title{\sf Light Nuclei near Neutron and Proton Drip Lines 
in the Relativistic Mean-Field Theory}
\author{G.A. Lalazissis$^{1,4}$, A.R. Farhan$^2$, M.M. Sharma$^{2,3}$,\\
$^1$Physik Department, Technische Universit\"at M\"unchen, Germany\\
$^2$Physics Department, Kuwait University, Kuwait 13060\\
$^3$Max Planck Institut f\"ur Astrophysik,\\
D-85740 Garching bei M\"unchen, Germany,\\
$^4$Department of Theoretical Physics, \\
Aristotle University of Thessaloniki, \\
Thessaloniki 54006, Greece}

\begin{document}
 
\maketitle
\begin{abstract}
We have made a detailed study of the ground-state properties of
nuclei in the light mass region with atomic numbers Z=10-22 in
the framework of the relativistic mean-field (RMF) theory.
The nonlinear $\sigma\omega$ model with scalar self-interaction
has been employed. The RMF calculations have been performed in an 
axially deformed configuration using the force NL-SH. We have 
considered nuclei about the stability line as well as those close 
to proton and neutron drip lines. It is shown that the RMF results 
provide a good agreement with the available empirical data. 
The RMF predictions also show a reasonably good agreement
with those of the mass models. It is observed that nuclei in this mass 
region are found to possess strong deformations and exhibit shape 
changes all along the isotopic chains. The phenomenon of the shape 
coexistence is found to persist near the  stability line as well as 
near the drip lines. It is shown that the magic number N=28 is quenched 
strongly, thus enabling the corresponding nuclei to assume strong 
deformations. Nuclei near the neutron and proton drip lines in this 
region are also shown to be strongly deformed. 

\end{abstract}

\newpage
\baselineskip=24pt

\section{\bf Introduction}
The advent of radioactive beams and emergence of several 
facilities to produce these has provided a possibility to study the 
structure and properties of nuclei far away from those known to us 
so far \cite{Rkl.92,MS.93,Tan.95,Ver.96,Proc.96}.
These so-called "exotic" nuclei transcend the valley of stability 
and possess the extreme ratios of protons to neutrons on 
both the sides. An increase in each unit of isospin makes  
nuclei vulnerable to decay and consequently nuclei in the 
extreme domains of the periodic table are far short-lived.
An access to these nuclei with a view to study their structure and
properties poses a serious experimental challenge. At the same time, 
a comprehension of the nuclear properties in such areas is essential. 
Such regions of the periodic table are now becoming far more accesible
to experimental studies \cite{Proc.96}. A knowledge of the structure 
of these nuclei is expected to facilitate the understanding of the 
processes responsible for synthesis of heavy elements.

Numerous theories have so far been applied to study structure of nuclei. 
The shell model provides the essential backbone of the most of the 
approaches which are based upon the treatment of a nucleus within 
a mean field. Some of these approaches are based upon mean-field 
generated from the density-dependent Skyrme functional within 
the Hartree-Fock approach \cite{Vau.73,QF.78}.
Both the zero-range and finite-range forces are used. 
The computational ease has made the zero-range Skyrme Hartree-Fock
approach a useful tool to discern properties of nuclei. The finite-range 
forces of the Gogny type has 
had only limited applications \cite{DGG.75,DG.80}
owing to the complicated handling of the associated 
functions and a longer computational time required. Both the zero-range
and the finite-range forces have achieved their successes.
Such studies have mostly been confined to nuclei closer to the stability
line. Efforts are on way to employ these approaches also to extreme
regions of nuclei. The role of the spin-orbit interactions and 
consequently the shell effects in regions far away from the 
stability line are recently being discussed. The shell effects
and their influence on nuclear properties about the drip line
is an important question which needs to be addressed.

The relativistic mean-field (RMF) theory \cite{SW.86,Rei.89,Ser.92,Ring.96}
has recently proven to be a powerful tool to describe and predict 
the properties of nuclei. The outstanding problem of the anomalous 
kinks in the isotope shifts of Pb nuclei \cite{TBF.93}, which remained
intractable for a long time, could be successfully resolved with 
in the RMF theory \cite{SLR.93}. However, in our studies with the 
RMF theory a few subtle differences have emerged vis-a-vis the Skyrme 
approach. It is realized that the spin-orbit interaction 
in the RMF theory, which arises as a result of 
the Dirac-Lorentz structure of the nuclear interaction leads to 
an implicit density dependence of the spin-orbit potential.
The issue of the isospin 
dependence of the spin-orbit interaction has also been 
raised recently \cite{LSK.94,SLK.95,RF.95}. An introduction of the isospin
dependence in the spin-orbit channel, derived in a fashion analogous
to the RMF theory seems to cure the problem of the density-dependent 
Skyrme approach \cite{SLK.95}. Consequently, this feature 
taken from the RMF theory is successful in obtaining the anomalous 
kink in the isotopic shifts of Pb nuclei using the Skyrme approach. 
It remains to be seen how this modification of the Skyrme 
theory will fare for nuclei away from the stability line.

The RMF theory has shown its success in reproducing isotope shifts 
and deformation properties of nuclei in the rare-earth region \cite{LS.96}.
The peculiar behaviour of the  empirical isotopic shifts \cite{Ott.89}
of rare-earth nuclei as well as those of the well-known experimental 
data on the chains of Sr and Kr nuclei could be described
successfully by Lalazissis and Sharma \cite{LS.95} for the first 
time. It is worth recalling that the Skyrme approach faced a 
difficulty in reproducing the isotope shifts of Sr nuclei, 
whereby the method of generator coordinates was employed to 
tackle this problem \cite{Bonche.91,Heen.93}. 
On the other hand, the RMF theory owes its success in providing a 
very good description of the isotopes shifts partly to better 
shell effects and in part to an improved description of the 
deformations of nuclei, which affect the isotope shifts accordingly.
The RMF theory has also been extended to many deformed nuclei 
earlier in ref. \cite{Werner.94,Werner.96,Ren.96} and it has been shown
that its predictions compare favourably with the known properties.
An extensive study of nuclei in the Sr region has been made recently 
in ref. \cite{Maharana.96}. Using the deformed RMF theory a fairly good
representation of the experimental data has been obtained.
In another approach based upon modification of meson masses with
density, properties of Ne and Mg isotopes have been described
using the deformed Hartree-Fock-Bogoliubov theory \cite{Grumm.96}.

Nuclei close to neutron drip-line present an interest due to
their relevance to r-process nucleosynthesis \cite{Kra.88}. 
Nuclei relevant to the neutron-rich nucleosynthesis lie typically in 
the region of Zr (about Z=36-44) and have an enormously large neutron 
excess.  Such nuclei are expected to remain inaccesible 
to experiments for a forseeable future. However, a knowledge of
these nuclei is necessary in order to understand the nuclear 
abundances \cite{Kra.93}. There exist several theoretical models 
and chiefly among them various mass models \cite{Hau.88}
which can predict the properties of nuclei close to the neutron
drip lines. These extrapolations are based naturally upon the
their respective Ansatz's which are fitted to obtain various
sets of available nuclear properties about the stability line.
The main emphasis is on a better reproduction of a large number of 
nuclear masses known over the periodic table. The Finite-Range 
Droplet Model (FRDM) \cite{MNM.95} and the Extended Thomas-Fermi 
with Strutinsky Integral (ETF-SI) \cite{APD.92}
are the best known mass models which are fitted exhaustively
to a large body of nuclear data. In both these mass 
formulae attempts have been made to obtain a best possible
description of nuclear masses and deformation properties.
It is, however, not clear whether it is possible to reproduce 
other nuclear properties such as isotope shifts and nuclear sizes 
within these models.

These mass models also extrapolate the properties of nuclei in the
region of drip lines. It has been observed that the FRDM and the 
ETF-SI predict the binding energies of nuclei close to the neutron 
drip lines, both of which are indicative of stronger shell effects. 
In the RMF theory, the shell effects about the neutron drip line have 
also been investigated.
It was shown by Sharma et al. \cite{SLH.94,SLHR.94} that the shell 
effects near the neutron drip line in Zr isotopic chains remain 
strong. This was found to be in consonance with that predicted
by the FRDM. In addition, many nuclei in this chain were observed to
be deformed and the deformations obtained in the RMF theory were
in very close agreement with those predicted by the FRDM. 
The relativistic mean-field theory has recently been extended 
by including the relativistic Hartree-Bogoliubov approach \cite{Meng.97} 
which takes the particle continumm properly into account. Again, the 
results indicate strong shell effects near the neutron drip line. 

Many nuclei close to the neutron drip line have been accessed 
experimentally (refer to the reviews
\cite{Rkl.92,MS.93,Tan.95,Ver.96,Proc.96} for more details). 
The drip line nuclei which are accessible to experiments at present 
lie in the low-mass region. Radioactive beams which could produce 
nuclei near drip line in the light-mass region are currently
available. Constant improvements are being made in the techniques
to produce nuclei far away from the stability line with a view to 
study the associated properties. In this paper we study the
properties of nuclei in the light-mass region in the RMF theory.
We have selected several isotopic chains from Z=10 to Z=24.
The nuclei included in these chains encompass both the neutron
and the proton drip lines. The calculations have been performed 
within an axially symmetric deformed configuration, the details
of which are provided in the section below. We will discuss
the results on the the nuclear sizes and nuclear deformations.
A comparison of our results will be made with the available 
experimental data. We will also compare our predictions for
the extreme regions with the predictions of the FRDM and ETF-SI,
where available. We will discuss the salient features of our
results in a broader perspective of the properties of nuclei 
about the drip lines. In the last section we provide a summary 
of our results.

\section{\bf Relativistic Mean-Field Theory}

The starting point of the RMF theory is a Lagrangian density \cite{SW.86}
where nucleons are described as Dirac spinors which interact via the
exchange of several mesons. The Lagrangian density can be written in
the following form:
\begin{equation}
\begin{array}{rl}
{\cal L} &=
\bar \psi (i\rlap{/}\partial -M) \psi +
\,{1\over2}\partial_\mu\sigma\partial^\mu\sigma-U(\sigma)
-{1\over4}\Omega_{\mu\nu}\Omega^{\mu\nu}+\\
\                                        \\
\ & {1\over2}m_\omega^2\omega_\mu\omega^\mu
-{1\over4}{\bf R}_{\mu\nu}{\bf R}^{\mu\nu} +
 {1\over2}m_{\rho}^{2}
 \mbox{\boldmath $\rho$}_{\mu}\mbox{\boldmath $\rho$}^{\mu}
-{1\over4}F_{\mu\nu}F^{\mu\nu} \\
\                              \\
\ &  g_{\sigma}\bar\psi \sigma \psi~
     -~g_{\omega}\bar\psi \rlap{/}\omega \psi~
     -~g_{\rho}  \bar\psi
      \rlap{/}\mbox{\boldmath $\rho$}

     \mbox{\boldmath $\tau$} \psi
     -~e \bar\psi \rlap{/}{\bf A} \psi
\end{array}
\end{equation}

The meson fields included are the isoscalar $\sigma$ meson, the 
isoscalar-vector $\omega$ meson and the isovector-vector $\rho$ meson. 
The latter provides the necessary isospin asymmetry. The bold-faced 
letters indicate the isovector quantities. The model contains also a 
non-linear scalar self-interaction of the $\sigma$ meson :

\begin{equation}
U(\sigma)~={1\over2}m_{\sigma}^{2} \sigma^{2}~+~
{1\over3}g_{2}\sigma^{3}~+~{1\over4}g_{3}\sigma^{4}
\end{equation}
The scalar potential (2) introduced by Boguta and Bodmer \cite{BB.77}
has been found to be necessary for an appropriate description 
of surface properties,  although several variations of the 
non-linear $\sigma$ and $\omega$ fields have recently been 
proposed \cite{Furn.96}. M, m$_{\sigma}$, m$_{\omega}$ and 
m$_{\rho}$ denote the nucleon-, the $\sigma$-, the $\omega$- and the 
$\rho$-meson masses respectively, while g$_{\sigma}$, g$_{\omega}$, 
g$_{\rho}$ and e$^2$/4$\pi$ = 1/137 are the corresponding coupling 
constants for the mesons and the photon.

The field tensors of the vector mesons and of the electromagnetic
field take the following form:
\begin{equation}
\begin {array}{rl}
\Omega^{\mu\nu} =& \partial^{\mu}\omega^{\nu}-\partial^{\nu}\omega^{\mu}\\
\          \\
{\bf R}^{\mu\nu} =& \partial^{\mu}
                  \mbox{\boldmath $\rho$}^{\nu}
                  -\partial^{\nu}
                  \mbox{\boldmath $\rho$}^{\mu}
                  -g_{\rho}(
                  \mbox{\boldmath $\rho$} \times
                  \mbox{\boldmath $\rho$})\\
\                \\
F^{\mu\nu} =& \partial^{\mu}{\bf A}^{\nu}-\partial^{\nu}{\bf A}^{\mu}
\end{array}
\end{equation}
The classical variational principle gives rise to the equations of motion. 
Our approach includes the time reversal and the charge conservation.
The Dirac equation can be written as:
\begin{equation}
\{ -i{\bf {\alpha}} \nabla + V({\bf r}) + \beta [ M + S({\bf r}) ] \}
~\psi_{i} = ~\epsilon_{i} \psi_{i}
\end{equation}
where $V({\bf r})$ represents the $vector$ potential:
\begin{equation}
V({\bf r}) = g_{\omega} \omega_{0}({\bf r}) + g_{\rho}\tau_{3} {\bf {\rho}}
_{0}({\bf r}) + e{1+\tau_{3} \over 2} {\bf A}_{0}({\bf r})
\end{equation}
and $S({\bf r})$ being the $scalar$ potential:
\begin{equation}
S({\bf r}) = g_{\sigma} \sigma({\bf r})
\end{equation}
the latter contributes to the effective mass as:
\begin{equation}
M^{\ast}({\bf r}) = M + S({\bf r})
\end{equation}
The Klein-Gordon equations for the meson fields are time-independent
inhomogenous equations with the nucleon densities as sources.
\begin{equation}
\begin{array}{ll}
\{ -\Delta + m_{\sigma}^{2} \}\sigma({\bf r})
 =& -g_{\sigma}\rho_{s}({\bf r})
-g_{2}\sigma^{2}({\bf r})-g_{3}^{3}({\bf r})\\
\         \\
\  \{ -\Delta + m_{\omega}^{2} \} \omega_{0}({\bf r})
=& g_{\omega}\rho_{v}({\bf r})\\
\                            \\
\  \{ -\Delta + m_{\rho}^{2} \}\rho_{0}({\bf r})
=& g_{\rho} \rho_{3}({\bf r})\\
\                           \\
\  -\Delta A_{0}({\bf r}) = e\rho_{c}({\bf r})
\end{array}
\end{equation}
The corresponding source terms are
\begin{equation}
\begin{array}{ll}
\rho_{s} =& \sum\limits_{i=1}^{A} \bar\psi_{i}~\psi_{i}\\
\             \\
\rho_{v} =& \sum\limits_{i=1}^{A} \psi^{+}_{i}~\psi_{i}\\
\             \\
\rho_{3} =& \sum\limits_{p=1}^{Z}\psi^{+}_{p}~\psi_{p}~-~
\sum\limits_{n=1}^{N} \psi^{+}_{n}~\psi_{n}\\
\                    \\
\ \rho_{c} =& \sum\limits_{p=1}^{Z} \psi^{+}_{p}~\psi_{p}
\end{array}
\end{equation}
where the sums are taken over the valence nucleons only. It should also
be noted that the present approach neglects the contributions of
negative-energy states ($no-sea$ approximation), i.e. the vacuum is not
polarized. The Dirac equation is solved using the oscillator expansion 
method \cite{GRT.90}.

\section{\bf Details of Calculations}

As most of the nuclei considered here are open shell nuclei, 
pairing has been included using the BCS formalism. We have used 
constant pairing gaps which are taken from the empirical particle 
separation energies of neighbouring nuclei. The centre-of-mass 
correction is taken into account by using the zero-point energy 
of a harmonic oscillator. For nuclei in the extreme regions, 
where nuclear masses are not known, we have used the prescription 
of M\"oller and Nix \cite{MN.92} to calculate the neutron
and proton pairing gaps. Accordingly,

$$
\Delta_n = {4.8\over N^{1/3}}\\
\Delta_p = {4.8\over Z^{1/3}}
$$

The number of shells taken into account is 12 for both the Fermionic and
Bosonic wavefunctions. It should be noted that for convergence reasons
14 shells were also considered. It turned out, however, that
the difference in the results is negligible and therefore all the
calculations reported in the present work were performed in a
12 shells harmonic-oscillator expansion.

In this paper we have used the force NL-SH. It has been shown that the 
force NL-SH provides excellent results \cite{SNR.93} for nuclei 
on both the sides of the stability line. The appropriate value of
the asymmetry energy of this force renders an adequate description of 
nuclei on both the neutron rich as well as on the proton rich sides. 
The parameters of the force NL-SH are:

\par\noindent
M = 939.0 MeV; $m_\sigma$ = 526.059 MeV; $m_\omega$ = 783.0 MeV;
$m_\rho$ = 763.0 MeV;
\par\noindent
$g_\sigma$ = 10.444; $g_\omega$ = 12.945; $g_\rho$ = 4.383;
$g_2$ = $-$6.9099 fm$^{-1}$ ; $g_3$ = $-$15.8337.

Here $g_2$ is in fm$^{-1}$ and $g_3$ is dimensionless.

\section{\bf Results and Discussion}

\subsection{Binding Energies}

The total ground-state binding energies of nuclei obtained in the RMF
calculations for various isotopic chains are shown in Tables 1 and 2. 
The results are obtained from the axially deformed RMF calculations 
with 12 oscillator shells using the force NL-SH. The binding 
energies from the mass models FRDM and ETF-SI are also shown for 
comparison wherever available. In the last column
of the tables the experimental binding energies 
\cite{AW.93} of the known isotopes are given. 
The binding energies in the RMF are in good agreement with
the empirical values. The quality of agreement is particularly good for
nuclei which are proton rich or neutron rich. For nuclei close to the
stability line a slight disagreement between the RMF predictions and
the experimental values appears. This disagreement is about 1-2\% for
some chains. It is a bit exaggerated for S nuclei. For the isotopic
chains of Ti and Cr the RMF binding energies show an overall
excellent agreement with the empirical data. A comparison of the RMF
predictions in the unknown region with those of the FRDM shows that
the two values show a general agreement for all the chains except
for S and Ar. Agreement between the RMF and FRDM predictions for
nuclei close to the proton and neutron drip lines is fair. Only for
S and Ar nuclei near the neutron drip line, the RMF shows a slightly
stronger binding than in the FRDM. 

The predictions of ETF-SI are available only for a few nuclei for
low Z chains such as Ne and Mg. For the other chains such as Si, S
and Ar, the ETF-SI results available for nuclei on the neutron-rich
side show that the level of agreement amongst the RMF, FRDM and ETF-SI
is about the same, i.e., the ETF-SI predictions agree reasonably well
with the RMF and FRDM. However, for the isotopic chains of Ti and Cr,
the ETF-SI tends to underestimate the binding energies of nuclei
close to the neutron drip line as compared to the RMF and the FRDM.

The RMF results on the binding energies of nuclei in the isotopic 
chains considered here show that the RMF theory is able to provide 
an overall description of the experimental binding energies. A good
representation of the experimental binding energies was also 
observed for deformed nuclei in the isotopic chains of 
Sr, Kr and Zr \cite{LS.95} as well as in many rare-earth 
chains in our earlier work \cite{LS.96}. The results for a broad 
range of isospin which includes both the neutron 
drip and proton drip lines, where experimental data is not known,
are consistent with the predictions of the FRDM. Thus, the RMF theory 
having only 6 effective parameters which are fitted only to a very limited
number of nuclei describes much of the experimental binding
energies consistently. 

\subsection{\bf Nuclear Radii and Sizes}

The $rms$ charge and neutron radii obtained from the deformed
relativistic mean-field calculations are shown in Figs. 1-2. 
The charge radii show a peculiar
behaviour that light isotopes of all the chains shown in the
above figures exhibit a higher charge radius as compared to their
heavier counterparts. The charge radii assume a minimum value
for some intermediate mass nuclei and it again increases for higher
masses. This behaviour is akin to an inverted parabola as a function 
of mass. Such a feature is demonstrated commonly by nuclei in 
deformed regions (see the review by Otten \cite{Ott.89} for details),
whereby nuclei a few neutron number below the most stable nucleus
possess a larger charge radius. This is especially exemplified by the 
Sr and Kr chains \cite{Buch.90,Keim.95}  
where the lighter isotopes have a bigger $rms$ charge 
radius than the heavier ones. However, an increase in the
charge radius for very neutron deficient nuclei stems from
the protons in the outermost orbitals. These protons are extended 
into space and create a proton skin. This skin is subdued considerably
due to the Coulomb barrier which tends to bind the protons more strongly.

The experimental charge radii \cite{deVries.87} of several
nuclei for each isotopic chain are shown by solid circles. 
It can be seen that the RMF predictions show a good agreement 
with the known experimental data. We also show in Figs. 1-2 the 
charge radii from the mass model ETF-SI for comparison. 
For Ne, Mg and Si only a few predictions from the ETF-SI are 
available. However, for the isotopic chains above S, 
ETF-SI values are available for the neutron-rich side. 
Above Ca, the predictions of the ETF-SI are are available for 
isotopes on both the sides of the valley of stability.
For nuclei close to the stability line, the ETF-SI results show
a fair agreement both with the experimental and the RMF values. However,
the ETF-SI results tend to overestimate the RMF results for the 
neutron-rich side, whereas for the proton-rich side the ETF-SI predictions
are smaller than the RMF predictions. This feature may be due to the
difference in the isospin dependence of the spin-orbit potential
in the RMF theory and the Skyrme Ansatz \cite{SLK.95}.

The neutron radii in Figs. 1-2 show an increasing trend  
with mass for all the isotopic chains. For highly neutron 
deficient (proton rich) nuclei the neutron $rms$ radius is much 
smaller than the corresponding charge radius. This has the obvious
implication that the neutron deficient nuclei possess an extended 
proton skin. This aspect will become clear in Figure 3 on the neutron 
skin thickness ($r_n - r_p$). 

The neutron radii exhibit a kink about the magic neutron numbers
N=20 and N=28. The kink at N=20 is prominent for Ne, Mg and Si isotopes. 
It becomes weaker for heavier isotopic chains of S, Ar, Ti and Cr. 
In these chains a moderate kink appears at N=28 due to its vicinity
to the neutron drip line. The end of each isotopic chain in Figs. 1-2 
depicts nuclei close to the neutron drip line. For these nuclei the 
neutron radii show a strong increase as a consequence of a large 
spatial extension of neutrons. This should imply an existence of 
neutron halos for nuclei close to the neutron drip line. 

The curves for the neutron radii cross those for the charge radii,
for all the isotopic chains, at a mass number which corresponds to
a neutron number higher than proton numbers by 2 for light chains
and about 4 for heavier chains. This difference can be explained
in part by the fact that the charge radii presented are not just
the proton radii but are convoluted for the finite size of protons.
Secondly, the Coulomb repulsion extends the mean field of protons 
slightly beyond that of the neutrons. 

For the isotopic chain of Ca, the neutron radii show an unusual
parabolic behaviour between A=40 (N=20) and A=48 (N=28). Since
the proton number for Ca isotopes is a strong magic number, it
seems to enforce a very strong N=28 magic shell thus leading to
a significant reduction in the radius of $^{48}$Ca as compared to
its lighter neighbours. This results in a strong kink in the 
neutron radius at N=28. Such a feature is not seen in other 
isotopic chains where the effect of the N=28 shell tends to 
diminish. 

Fig. 3 shows the neutron skin thickness $(r_n - r_p)$ obtained from
the RMF calculations for the isotopic chains from Z=10 to Z=24.
All the isotopic chains considered here encompass nuclei close to 
the proton drip line as well as those close to the neutron drip line. 
For highly neutron deficient nuclides i.e. close to proton drip, the 
neutron skin thickness is negative. This symbolizes a proton 
skin in all these nuclei. On the other side i.e. for nuclei possessing a
large number of neutron excess and which are close to the neutron drip 
line, a positive and large neutron skin can be easily
seen. This is suggestive of a significant neutron halo in
the extremely neutron-rich nuclei. 

The figures show an interesting point about the difference in the
magnitude of the proton and neutron skins. For a given proton excess
i.e. (Z-N), the corresponding proton skin thickness is seen 
be much larger than the neutron skin thickness for the same value of 
neutron excess (N-Z). This is obviously due to the Coulomb repulsion 
of protons. The Coulomb effect will also have a consequence that
the proton drip line is expected to arise much faster in going away from
the stability line than the corresponding neutron drip line. 

The effect of shell closure in neutron skin-thickness can be seen 
easily in some nuclear chains. The prominent case of such an effect 
is visible in the Ca isotopes where strong kinks at N=20 and N=28 
are present. In other isotopic chains such as with the lower atomic 
numbers (upper panel) the kinks at N=20 and N=28 are also to be seen. 
In these cases, the shell N=20 is strongly magic and some effect of 
N=28 is still present. In contrast, for nuclei in the lower panel, 
i.e., above S, the kink at N=28 becomes minimal. This is attributed to
a diminishing shell gap at N=28 in these nuclei.

We show the neutron and proton density distributions of a few neutron 
rich isotopes in Fig. 4. The nuclei included are $^{42}$Mg, $^{56}$S and
$^{80}$Cr. All these nuclei are close to the predicted neutron drip 
line. Here the neutron density is in stark contrast to the corresponding
proton density. The density of neutrons in the interior of the nuclei
is larger as compared to the proton density and it extends considerably
in space as compared to the proton density. This difference in the 
density distributions is a reminder of the possibility 
of a large neutron halo in nuclei near the neutron drip line.

In Fig. 5 we show the density distributions for a few nuclei
near the proton drip line. We have selected the nuclei 
$^{20}$Mg, $^{26}$S and $^{40}$Cr to demonstrate the effect of the
proton-excess $(Z-N)$ on the density distributions. The central density 
of protons as well as the proton density at the surface, both exceed
the corresponding neutron densities. The spatial extension of the 
proton density can also be seen in all the three cases. Examining the
$S_{2p}$ values of these, it can be  said that all these nuclei are in 
the vicinity of the proton drip line. In particular, the proton-excess
of $^{20}$Mg and $^{26}$S is only 4 and 6, respectively. Thus, due to
the strong Coulomb repulsion only a moderate value of the proton-excess
seems to suffice to create nuclei close to the proton drip line.
This can be contrasted strongly with the nuclei near the neutron
drip-line. The neutron-excess $(N-Z)$ required to produce the Mg and 
S nuclei close to the neutron-drip line amounts to about 16 and 24, 
respectively. These numbers are about a factor of 4 larger than the 
corresponding numbers for the proton-drip line. This difference
demonstrates the strength of the Coulomb force which causes
the proton drip line nuclei to appear much closer to the stability
line than the neutron drip line. An obvious consequence of this effect
is the constraint on the magnitude or on the relative extension of the
proton density vis-a-vis the neutron density. Thus, the proton halo 
in such nuclei is reduced considerably due to low values of the 
proton-excess.

\subsection{The Separation Energies}

The 2-neutron separation energies $S_{2n}$ for all the isotopic chains
are shown in Fig. 6. Each curve represents an isotopic chain. 
The 2-neutron separation energies for isotopic chains with a higher 
Z number start from a higher value. The $S_{2n}$ values show the 
charateristic decrease with an increase in neutron number. 
For a reasonably large neutron number each curve approaches the 
vanishingly low $S_{2n}$ values. This represents the onset of the
neutron drip. Inevitably, the $S_{2n}$ curves for low-Z chains
are rather short for the neutron-drip line to reach. 
This would make many of the drip nuclei accessible to future 
experiments. On the other hand, the curves for high-Z isotopic chains 
become longer until the $S_{2n}$ values approach a nearly vanishing 
value for an access to the neutron drip line. Thus, nuclei near
the neutron drip line possess a large neutron excess.

In Fig. 7 we show the 2-proton separation energies $S_{2p}$. The curves
are labelled by a neutron number which corresponds to a given isotonic
chain. The lowest curve is for N=8 and the highest one denoted by
the dashed line is for N=38. The $S_{2p}$ values show an usual decreasing 
trend as the proton number Z increases. For nuclides with low neutron
number N, the $S_{2p}$ curve approaches a vanishing value rather fast.
This implies that the proton drip line in low N nuclides can be accessed
more easily than would be the case for high N nuclides. For nuclides
with higher number of neutrons, a relatively larger number of protons
are needed for the proton drip. Translating this fact into the proton
drip line for each element (Z), it can be seen that for a given Z, the 
point at which a $S_{2p}$ curve reaches about zero value, gives its
associated neutron number. It can be realized that the asymmetry between
Z and N for proton drip is small for low Z elements and that this
asymmetry increases only moderately for high Z elements. This is in contrast
to the neutron drip line which occurs for large values of neutron excess.
This point has also been elaborated above in Figs. 4 and 5.

\subsection{The Quadrupole Deformation } 

Nuclei in all the isotopic chains, except Ca, considered in this study 
are prone to being deformed. The quadrupole deformation $\beta_2$ 
obtained from the axially deformed mean-field minimization are shown
in Figs. 9 and 10. The numerical values of the quadrupole
deformation $\beta_2$ and the hexadecapole deformation $\beta_4$
as obtained in the RMF theory with the force NL-SH are given in 
Tables 3-5. It is observed that some nuclei also show a secondary 
minimum in the energy. The corresponding deformation for the secondary
minimum is given within paretheses. The results show that a large number 
of nuclei in these chains are highly deformed. Nuclei with prolate
as well as oblate deformations are predicted.

Nuclei in the Ne and Mg chains are overwhelmingly prolate. In all the 
chains shown in Fig. 9, nuclei with N=8 are manifestly spherical. 
It shows that the magic number N=8 enforces a spherical shape on the 
mean-field irrespective of the associated proton number. Going away 
from N=8, nuclei in the Ne and Mg chain assume a highly prolate shape
with $\beta_2$ close to 0.40. As the shell closure N=20 is approached,
nuclei tend to sphericise. This effect can be seen in all the
isotopic chains shown in Figs. 9 and 10 including $^{32}$Mg and $^{36}$S 
which are predicted to be spherical. Above N=20, the Ne and Mg nuclei
take up highly prolate shapes in the RMF theory.
In comparison, the Si isotopes undergo a sequence of shape 
transitions spherical-prolate-oblate-spherical and again 
spherical-prolate-oblate in going from A=22 (N=8) to A=46 (N=32).
On the other hand, in the S and Ar chains nuclei are changing 
shape between prolate and oblate at regular intervals in the neutron
number. The highly neutron-rich nuclei close to the neutron-drip
line are shown to take a well deformed oblate shape for Si, S and Ar.

In the chains of Si, S and Ar, it is noticed that the shape transition
from a prolate to an oblate shape is shown to occur from N=12 
to N=14. Thus, an addition of a pair of neutrons to N=12 tends to
switch the shape of a nucleus from prolate to oblate. Such a
transition is not shown for Ne and Mg isotopes. However, in these 
nuclei a well-deformed oblate shape which coexists with the 
prolate shape, is predicted at N=14 in the RMF theory.

The S and Ar isotopes show several of prolate to oblate transitions.
In contrast, Ti and Cr chains which have proton numbers above the strong 
magic number Z=20, are free from such prolate to oblate transitions except 
for a slightly oblate shape in the vicinity of N=36-38. On the whole, Ti and 
Cr nuclides are predominantly prolate with a succession of a few shape
changes to a spherical one. The transitions to a spherical shape from 
a prolate one at regular intervals and the behaviour akin to a brigde
is noteworthy for the Cr nuclei. 

\subsubsection{Comparison with the Experimental Data}

It is well known that nuclei in this region are very strongly
deformed. Especially, in the low-Z chains, nuclei assume shapes
of unusual proportion. We compare the RMF predictions on the
quadrupole deformation ($\beta_2$) of nuclei with the available 
experimental data in Table 6. The experimental data have been
taken from the compilation of $\beta_2$ by Raman et al. \cite{Ram.87}
from the analysis of the BE(2) values. It is noticed from Table 6
that a few Ne nuclei take up a very large value of $\beta_2$. Such a 
magnitude is inexplicable in the RMF theory and particularly within
an axially deformed configuration we have adopted. Some of these
nuclei are hypothesized to possess a triaxial shape in other models.
We can not account for the triaxiallity in the present work.
On the other hand, for the Ne isotopes the RMF theory produces 
from moderate to strong deformations as in the traditional
parlance. For $^{20}$Ne, the RMF deformation is low as compared to
the experimental value. This difference is due to the complex nature 
of the potential-energy landscape for $^{20}$Ne. The RMF theory,
however, comes close to the empirical value for $^{22}$Ne. 

The quadrupole deformations of several Mg isotopes are described
well in the RMF theory. The RMF theory underestimates the empirical
values slightly. This is also the case for the Si isotopes. A few Si 
isotopes are predicted to be oblate in the RMF theory. A comparison 
of the RMF predictions with the data is not straightforward
as the empirical values do not give the sign of the deformation.

A notable difference in the deformation appears for $^{32}$Si.
It is predicted to be nearly spherical in the RMF theory, whereas
the empirical data show that it has a $\beta_2 = 0.34$. The spherical
nature of $^{32}$Si in the RMF can be understood in that that the
associated neutron number N=18 lies very close to the magic number
N=20 which overpowers the shape. A similar instance is seen for 
$^{40}$Ar which appears as spherical in the RMF theory.
All nuclei with N=20, such as $^{36}$S and $^{38}$Ar are explictly
spherical in the mean field. 

Recently, $\beta_2$ values for a few neutron-rich S and Ar isotopes 
have been inferred from the BE(2) measurements \cite{Scheit.96}.
The nuclei $^{38,40,42}$S and $^{44,46}$Ar have been found to possess
significant quadrupole deformations. The corresponding experimental
values are shown in Table 6. It can be seen that the RMF theory
provides $\beta_2$ values which are very close to the experimental 
ones for the nuclei $^{40}$S, $^{42}$S and $^{44}$Ar. However,
The nuclei $^{38}$S and $^{46}$Ar come out as spherical in the
RMF theory, whereas experimentally these nuclei are shown to be
reasonably deformed. On the other hand, for several other Ar isotopes,
the RMF deformations are close to the empirical values. The RMF values
also approach the empirical values for several other S isotopes.
 
Most of the Ti isotopes are predicted to be spherical in the RMF theory 
because of a strong influence from the magic number Z=20. However,
the Cr nuclei which move away from the influence of Z=20 are duly
deformed in the RMF theory and there is a reasonably good agreement 
with the empirical values. 

\subsubsection{Comparison with Mass Models}

The predictions for quadrupole deformation $\beta_2$ from the mass
models FRDM and ETF-SI are also shown for comparison in Figs. 9 and 10.
Results from ETF-SI are not available for Ne and Mg chains and are 
available only partly for the Si and S chains. For Ar, Ti and Cr chains
predictions of both the mass models have been compared with the RMF 
results in Fig. 10.

The FRDM results for Ne, Mg, Si and S chains represent roughly the
pattern predicted by the RMF theory. The FRDM predicts several oblate
shape transitions for Ne and Mg nuclei. Taking into account
the shape-coexistence in Ne and Mg isotopes, the RMF results would 
come close to the FRDM predictions. For Si and S chains, the trend of
the FRDM results seems to agree with that of the RMF results.
The differences, however, can be seen for very light (A=24-26) Si
isotopes which are predicted to be prolate in the RMF theory and 
shown to be oblate in the FRDM. Another difference in the RMF and
FRDM predictions is for S nuclei (A=54-56) close to the neutron drip 
line. Here, the shape of the nuclei is predicted to be close to
spherical in the RMF theory, whereas the FRDM predicts a highly
deformed prolate shape. The ETF-SI results on Si and S show many 
similarities as well as a few differences with the RMF and FRDM 
results. For example, S isotopes with A=46-48 are predicted to be 
prolate in the RMF and FRDM, whereas in the ETF-SI, these nuclei take 
up a well-deformed oblate shape. 

In the RMF theory, the Ar nuclei change the shape between a prolate
and an oblate one along the chain, whereas the nuclei in approaching 
the neutron drip-line assume a predominantly oblate shape. 
The FRDM results, on the other hand, predict that the 
low-mass as well as heavy-mass Ar isotopes are oblate
and that nuclei in the middle of the chain take up spherical 
shape. In comparison, the ETF-SI results, follow largely the 
trend predicted by the RMF theory. 

The Ti nuclei show a slightly different behaviour in the deformation
properties. A large number of Ti nuclei are seen to be spherical.
The vicinity of the proton number to the strongly magic number 
Z=20 lends a spherical shape to many of the Ti isotopes. 
Some of the Ti isotopes are, however, deformed with a nominal 
deformation $\beta_2$ between 0.10-0.15. Only the very light 
$^{34}$Ti nucleus is predicted to have $\beta_2$ about 0.20. 
A comparison of the RMF results with the FRDM shows that both the 
RMF and FRDM predict a prolate deformation for $^{34}$Ti, although 
in FRDM the $\beta_2$ value predicted for this nucleus is a factor 
of 2 more than that in the RMF. 

There is at least one Ti nucleus which takes an oblate shape. 
The nucleus $^{58}$Ti (N=36) has an oblate deformation 
$\beta_2 \sim -0.10$. Both the RMF and the FRDM predict an oblate 
shape for $^{58}$Ti with a very good agreement in the $\beta_2$ values.
For most of the other Ti isotopes, the FRDM predicts a predominantly 
spherical shape, which is in good agreement with the RMF results. 
The shape transition spherical-prolate-spherical from A=66 to A=72 
in the RMF theory appears very closely also in the FRDM. 
Thus, there are various similarities in the predictions of the 
RMF theory and the FRDM. The ETF-SI, on the
other hand, predicts that most of the Ti isotopes should have a
reasonably strong prolate deformation. This is at variance with the
predictions of the RMF and the FRDM. 

The $\beta_2$ values for the Cr isotopes show a most remarkable
feature in Fig. 10. In the RMF theory, the shape of the Cr nuclei
changes between a spherical and prolate one rather periodically. 
The cusps representing such a behaviour in the RMF theory are partly
reproduced in the ETF-SI and partly in the FRDM. The very light Cr
isotopes (A=38-40) close to the proton drip line are predicted to
assume a highly deformed prolate shape ($\beta_2 \sim 0.25-0.30$).
Both the FRDM and ETF-SI also predict similar deformations. However,
on adding a few pairs of neutrons to these nuclei, the FRDM shows a
spherical shape for nuclei up to A=52. The ETF-SI results, on the
other hand, show the undulations in the $\beta_2$ values upto A=60,
similar to the RMF predictions, however, without going to a spherical
shape at A=44 and A=52 as against the RMF where these nuclei take up
a spherical shape. Above A=60, there is a broad agreement between the
RMF and the FRDM results, whereas the ETF-SI results go out of tune
with the other results and predict highly prolate shapes between
A=60-70. In this region, the RMF and the FRDM results show a shape
transition prolate-spherical-prolate where the $\beta_2$ value reaches
the maximum (A=70) of the cusp. At A=74 and above, the RMF theory as
well as the mass models all predict a similar behaviour. The nucleus
$^{74}$Cr (N=50) is predicted to be spherical in the RMF theory and is
nearly spherical in the FRDM and ETF-SI. Nuclei above A=74 are close
to the neutron drip line. These nuclei (A=78-80) are predicted to have a
reasonably well-deformed prolate shape in the RMF. This is supported
by the mass models. Thus, for the Cr chain, nuclei near the proton
drip line as well as those near the neutron drip line are predicted 
to be prolate shaped in all the three results presented in Fig. 10.

\subsection{The Hexadecapole Deformation}

The hexadecapole deformation $\beta_4$ obtained in the RMF theory
are shown in Tables 3-5. It is as per common expectation that
$\beta_4$ is zero for nuclei with spherical shapes (magic numbers).
This is true for most magic numbers except for N=28. For nuclei with
N=28, we observe that the $\beta_4$ has a significant value for 
the Ne, Mg, Si and S isotopes as these nuclei possess a correspondingly
large quadrupole deformation $\beta_2$. For Ar, Ti and Cr nuclei with
N=28, however, $\beta_4$ is close to zero as these nuclei are 
obtained to be spherical under the influence of the vicinity of Z=20. 
Clearly, $\beta_4$ tends to vanish for all nuclei with a vanishing 
$\beta_2$.

Many nuclei in this region show strong hexadecapole deformation as
given in Tables 3-5. In the Ne chain, most nuclei have a zero or a
very small $\beta_4$. Here only the nucleus $^{34}$Ne is predicted
to have a large and positive $\beta_4$. For the case of Mg isotopes,
the $\beta_4$ values predicted for most nuclei are also small or
very moderate. In comparison, the FRDM predicts substantial
hexadecapole deformation for several nuclei in the Ne and Mg chains.
It can be noticed that $\beta_4$ values for many strongly deformed
(large $\beta_2$) nuclei in the RMF are small. On the other hand, 
the $\beta_4$ values in FRDM are proportionately 
larger for highly deformed nuclei. 
This situation with regard to $\beta_4$ persists also for Si isotopes
in the RMF theory vis-a-vis FRDM, with only an occasionally large
$\beta_4$ in RMF. For S isotopes, on the other hand, both the RMF
and FRDM predict moderate values for $\beta_4$, with FRDM predicting
a significantly large  $\beta_4$ only for a very few nuclei. 
As to the signs of the deformations, there seems to be a little 
correlation between the signs of $\beta_2$ and $\beta_4$ for nuclei 
in the Ne, Mg, Si and S chains. However, most S isotopes show a 
negative $\beta_4$ values in the RMF as well as in the FRDM. 

The RMF theory predicts small $\beta_4$ values for Ar, Ti and Cr
nuclei as well. For the Ar chain, in comparison, the FRDM predicts
strong hexadecapole deformation for many nuclei. A strong correlation 
between $\beta_2$ and $\beta_4$ is found for Ar nuclei.
Both the RMF and FRDM predict a negative $\beta_4$ for oblate
nuclei. This is the case for neutron deficient as well as for
neutron rich nuclei. As can be seen, most Ar nuclei are oblate
deformed. 

As one goes above Z=20, both the RMF and FRDM show small values
of $\beta_4$ for Ti and Cr chains. The FRDM, in addition, predicts
a reasonably large $\beta_4$ for a few Ti and Cr nuclei at the
beginning or at the end of the chains. For other nuclei, there
is a broad agreement between the RMF and FRDM on small values
of the hexadecapole deformation for these isotopic chains.

The ETF-SI predictions for $\beta_4$ are also shown in Tables 3-5.
The ETF-SI values for $\beta_4$ for the lighter chains are available
only scantily. For heavier chains, the ETF-SI predictions are known
for many nuclei. It may be remarked that by and large the ETF-SI
values, where available, are in line with the RMF predictions of 
small $\beta_4$. 

\subsection{Shell Closures and Deformations}

The nuclear structure and the ensuing shell effects associated
to the magic numbers are important to understanding the relative
abundances of nuclei. Nuclei are formed either through the
neutron-rich nucleosynthesis or via rapid neutron capture
(r-process). The existence of a major shell in the path of
the neutron-rich nucleosynthesis or r-process provides a 
stability to the nuclei concerned and thus influences the 
abundance of the product nuclei. For example, the shell effects 
around Ca affect the synthesis of nuclei and their abundances 
in the region of iron. 

The shell closures encountered in the mass range we have studied 
appear at the particle numbers 8, 20, 28 and 50. Here we examine 
the predictions of the RMF theory along with those of the FRDM 
and ETF-SI for deformations at and in the vicinity of these 
magic numbers.

Some nuclei such as $^{18}$Ne, $^{20}$Mg and $^{22}$Si included
in Fig. 9 have the neutron number N=8. Most of these nuclei are highly
deficient in neutrons and are close to the proton drip line. The
results of the RMF theory on $\beta_2$ show that all the above nuclei
are predicted to be spherical. Thus, the magicity of N=8 is retained
also about the proton drip line. The corresponding results from
the FRDM show that although the trend of the deformations of nuclei
with neutron numbers above N=8 are similar in the FRDM to those from
RMF with a few exceptions, the FRDM is predicting a marginally
prolate deformed nuclei for N=8. It is, however, not known how the
ETF-SI predictions for the N=8 magic number are ?

The next magic number N=20 is considered to be a prominently strong
shell closure. Calcium nuclei with Z=20 are well-known to be
strongly magic. All the isotopic chains considered here include
nuclei with N=20. It can be seen from Figs. 9 and 10 that nuclei
with N=20 in all the isotopic chains are predicted to be spherical
in the RMF theory. Thus, the strong shell closure at N=20 sphericizes
the total mean field. This feature is also to be seen in the
predictions of the FRDM for all the chains from Ne to Cr. The ETF-SI
results for N=20 nuclides for Ne, Mg and Si are not available. 
The ETF-SI predictions for $^{36}$S, $^{38}$Ar and $^{42}$Ti (N=20) 
show that these nuclei are spherical. 

Next, we consider properties of nuclei with the magic number
28. The nuclei with N=28 included in Figs. 9 and 10 are $^{40}$Mg, 
$^{42}$Si, $^{44}$S, $^{46}$Ar, $^{50}$Ti and $^{52}$Cr. The RMF
results show that the nuclei $^{40}$Mg ($\beta_2 = 0.45$),
$^{42}$Si ($\beta_2 = -0.34$) and $^{44}$S ($\beta_2 = -0.20$) 
are highly deformed. The nucleus $^{44}$S is oblate deformed in
its lowest energy state in the RMF theory. However, for this nucleus
there is a shape-coexistence of the oblate shape with a highly prolate 
shape ($\beta_2 = 0.38$) which lies about 30 keV above. These two
states are practically degenerate. Earlier, it was shown in 
\cite{Sor.93} that the shape and deformation of
a nucleus is very important to obtaining a correct $\beta$-decay
probability. On the basis of the measured decay half-lives, it was
surmised \cite{Sor.93} that the ground-state of the nucleus
$^{44}$S is oblate deformed. The oblate ground state for $^{44}$S
in the RMF theory is consistent with this conclusion. However,
it is not clear how much will be the influence of the prolate 
state on the $\beta$-decay probability.

The above examples serve to demonstrate that the shell
closure N=28 is very weak and that shell effects due to N=28 are 
quenched. However, this may not be a general feature as the nuclei
$^{46}$Ar, $^{50}$Ti and $^{52}$Cr maintain a spherical shape. This
may be due to a combined effect of the proton number, which is close
to the magic number Z=20, and that of the shell closure at N=28, which 
lends a spherical symmetry to the mean field. Thus, the number N=28
is not universally magic. Considering the results of the mass models,
the FRDM shows a behaviour similar to the RMF theory in the quenching 
of the shell effects at N=28 for Mg, Si and S nuclei. For the nuclei
Ar, Ti and Cr (with N=28) the FRDM shows that these nuclei take a
spherical shape similar to that predicted by the RMF. Thus, for these
nuclei the magicity of N=28 is maintained also in the FRDM.

The ETF-SI predictions for N=28 nuclei are available for isotopic 
chains heavier than Mg. It can be seen that the ETF-SI predicts
a well-deformed shape for $^{42}$Si and $^{44}$S. This is in good
accord with the predictions of the RMF and FRDM. However,
the ETF-SI predicts a deformed shape also for $^{46}$Ar, $^{50}$Ti
and $^{52}$Cr, which are all predicted to be spherical both in the RMF 
and FRDM. Thus, the ETF-SI maintains the quenching of the shell
effects and leads to a considerable weakening of the magicity of 
N=28 in all the above nuclei. The results of the ETF-SI for N=28 
nuclei are in a slight contrast with those of the RMF and the FRDM. 

Shell closure at N=50 has shown itself to be strong in the known
nuclei. However, when it is stretched to the extreme regions,
it is not known experimentally whether the magicity of N=50
continues to persist. The only isotopic chains where nuclei
with N=50 could be constructed are those of Ti and Cr.
The nuclei $^{72}$Ti and $^{74}$Cr with N=50 are expected to lie
close to the neutron drip line. It can seen from Fig. 10 that
Ti and Cr nuclei immediately below N=50 are predicted to be 
prolate deformed in the RMF theory, the deformation being stronger 
for Cr nuclei. The Cr nuclei above N=50 are also shown to take
a prolate shape. However, the Ti and Cr isotopes with N=50 do take 
up a spherical shape. Thus, the shell closure and the magicity of
N=50 near the neutron drip line is strong in the RMF theory. 
Both the FRDM and ETF-SI also predict a nearly spherical 
shape for nuclei with N=50 near the neutron drip line, thus
maintaining strong shell effects at N=50.

\subsection{Neutron-Proton Deformations}

As we have included long chains of several nuclei which 
encompass both the proton and neutron drip lines, the range of the
variation in the isospin of the nuclei is considerably large.
This provides a testing ground for the relative variation in the 
neutron mean-field vis-a-vis the proton mean-field as a function of 
the isospin. The position of both the proton and the neutron number 
of a nucleus from a closed shell is decisive to what shape the 
respective mean fields assume. From this point of view, it is 
illustrative to compare the deformations of the neutron and proton
mean-fields as a function of the isospin. In Figs. 11 and 12 we show 
the difference $\beta_2(n) - \beta_2(p)$ in the quadrupole 
deformations of the neutron and proton fields. 

For most of the Ne nuclei below A=26, which are deformed (see Fig. 9),
the difference in the deformation of the neutron and proton densities
is very small and close to zero except for the nucleus $^{22}$Ne which
shows that the neutron mean-field is slightly more deformed than
the proton mean field. Here the neutron number N=14, which lies in the
middle of N=8 and N=20, seems to produce more deformation for
neutrons. Nuclei between A=26 and A=32 are spherical and hence the
obvious value (zero) of $\beta_2(n) - \beta_2(p)$. An interesting 
aspect appears above A=32. The nuclei from A=34 to A=38 take up an
increasingly larger $\beta_2$ as the neutron number increases
(Fig. 9). It can be seen from Fig. 11 that for these nuclei, the
difference $\beta_2(n) - \beta_2(p)$ also increases with an increase
in the $\beta_2$ value with a successive addition of a pair of neutrons.
Thus, the additional neutrons to N=20 core tend to drive the neutron
matter to highly deformed  shape as compared to the proton matter.
This can be exemplified by stating that $^{38}$Ne, which shows a large 
deformation $\beta_2$= 0.44, has $\beta_2 (n) = 0.49$ and 
$\beta_2 (p) = 0.32$. Thus, there is a substantial difference between 
the deformations of the neutron and proton mean-fields. 

Isotopes in the Mg chain, except $^{32}$Mg (N=20), have been shown to 
possess a prolate shape in the RMF theory. The difference $\beta_2(n)
- \beta_2(p)$ is, however, shown to have a rather strange feature.
The light Mg isotopes below N=20 (A=32) show a negative  
$\beta_2(n) - \beta_2(p)$ and those above N=20 
have a positive $\beta_2(n) - \beta_2(p)$. This means that the 
light Mg isotopes with mass upto A=30 (N=18) have a more 
strongly deformed proton mean-field than the neutron
mean-field. For example, the highly deformed nucleus $^{22}$Mg,
which has a prolate deformation $\beta_2 = 0.43$,
has $\beta_2 (p) = 0.46$ and $\beta_2 (n) = 0.39$. 
Thus, proton number Z=12 produces a stronger
deformation than the neutron number N=10. For $^{24}$Mg, which has
an equal number of protons and neutrons, such a big difference in the
deformations in not apparent. However, for the nuclei $^{26}$Mg (N=14)
and $^{30}$Mg (N=18), the difference in $\beta_2(n) - \beta_2(p)$
is not insignificant and the deformation $\beta_2 (p)$ continues to be
bigger than the corresponding $\beta_2 (n)$. This shows that below
N=18 the deformation tendencies of Z=12 are far stronger than those of
any neutron number. The above effect is, however, taken over by the
neutrons which play a leading role in deforming the nuclei beyond
$^{32}$Mg (N=20), noting that all nuclei above and including $^{34}$Mg
are highly deformed ($\beta_2 \sim 0.4$). One observes a positive 
$\beta_2(n) - \beta_2(p)$ for nuclei. This difference increases to
a considerable value of about 0.09 ($\beta_2 (n) = 0.48$ and
$\beta_2 (p) = 0.39$) for $^{40}$Mg (N=28). This shows that although 
the neutron number N=28 constitues the known shell-closure, the 
neutron mean-field take up a very large deformation and forces the
nucleus to take a highly deformed shape ($\beta_2 = 0.45$). Thus,
the shell-closure at N=28 is washed out and the shell effects are
considerably weakened. 

The Si nuclei from A=22 (N=8) to A=36 (N=22) do not show any
marked difference in the deformations of neutrons and protons,
though Si nuclei with large prolate as well as oblate deformations
have been predicted in the RMF theory (see Fig. 9). However, for
isotopes including and above A=38 (N=24), $\beta_2 (n)$ becomes
larger than $\beta_2 (p)$. Such a behaviour persists even across
the shape transition from prolate to oblate at A=42 (Fig. 9) and
now the negative $\beta_2 (n)$ values are bigger than the negative 
$\beta_2 (p)$ values and hence a negative  $\beta_2 (n) - \beta_2 (p)$
as shown in Fig. 11. Thus, for heavier Si isotopes, a dominance of 
neutron deformation over the proton deformation is seen both for
the prolate and the oblate nuclei, as opposed to the light Mg 
isotopes where the proton deformation overweighed against the 
neutron deformation in the RMF theory. 

The behaviour of S isotopes towards the difference 
$\beta_2 (n) - \beta_2 (p)$ is similar to that of Si isotopes.
Although the S nuclei show a variety of shape transitions throughout
the isotopic chain, a closer look at the $\beta_2 (n) - \beta_2 (p)$
values shows that for $^{34}$S (N=18) the $\beta_2 (p)$ is bigger than
$\beta_2 (n)$. This is due to the vicinity of N=18 to the magic number
N=20, which enforces a less deformed neutron mean-field as compared
to the proton mean-field. All the other S isotopes have, however, 
a higher neutron deformation than the proton deformation. This is the 
case for both the prolate and oblate shapes in the lowest minimum. 

The case of $^{44}$S needs a special mention. This nucleus (N=28) has 
an oblate shape ($\beta_2 = -0.20$) in the lowest minimum. However, 
a highly deformed prolate shape ($\beta_2 = 0.38$) for this nucleus 
is almost degenerate (only 27 keV higher) with the oblate one.
The neutron-proton difference in the deformation for the oblate shape 
is marginally low and is shown in Fig. 11. However, for the prolate 
shape a large difference $\beta_2(n) - \beta_2(p) = 0.10$ 
($\beta_2(n) = 0.41$ and $\beta_2(p) = 0.31$) is observed.
It is interesting to note that notewithstanding the N=28 magic
number, the neutron mean-field is susceptible (softer) to the 
stronger deformation than the proton mean-field. 

The deformation curve ($\beta_2$) for Ar isotopes in the RMF theory as 
shown in Fig. 10 seems to be reflected considerably in the difference
$\beta_2 (n) - \beta_2 (p)$ as shown in Fig. 12. Only for the nuclei
$^{32}$Ar and $^{36}$Ar which show an oblate deformation (negative 
$\beta_2$), the difference $\beta_2 (n) - \beta_2 (p)$ is shown to
be positive. This implies that for all the deformed Ar isotopes except
these two nuclides, the neutron deformation $\beta_2 (n)$ exceeds the 
proton deformation $\beta_2 (p)$. This is the case also for the oblate
shaped nuclei. For example, $^{52}$Ar (N=34) which has an oblate
deformation $\beta_2 = -0.22$ has a $\beta_2 (n) = -0.24$ and
$\beta_2 (p) = -0.19$, with a substantial difference of
$\beta_2 (n) - \beta_2 (p) \sim -0.05$. Thus, also for the Ar isotopes,
the neutron mean-field is more succeptible to deformation than the
proton mean-field. This is due to the reason that the proton number
Z=18 being very close to the magic number Z=20 lends itself to
the spherical tendency. 

The Ti isotopes also replicate the behaviour of their $\beta_2$
values (Fig. 10) in the difference $\beta_2 (n) - \beta_2 (p)$ as
shown in Fig. 12, i.e. nuclei with a positive $\beta_2$ have a positive
difference $\beta_2 (n) - \beta_2 (p)$ and those with a negative
$\beta_2$ have a negative $\beta_2 (n) - \beta_2 (p)$.
The only exception is the nucleus ${34}$Ti (N=12) which has a
quadrupole deformation $\beta_2 \sim 0.21$ and the difference in the
deformations of its neutron and proton densities is close to zero.
For other deformed Ti nuclei, the differences $\beta_2 (n) -
\beta_2(p)$ are proportional to the corresponding $\beta_2$ values.
However, the differences in the deformations of neutron and proton
densities are moderate as compared to those found for some Ne and Mg
isotopes and are found to be slightly smaller than those for some Si,
S and Ar nuclei. This is due to the fact that Ti has a proton number
Z=22 which is close to the strongly magic number Z=20. Thus, the Ti
isotopes oppose tendencies which would produce vigourous changes and
differences in the neutron and proton deformations. 

The behaviour of Cr isotopes for the $\beta_2 (n) - \beta_2(p)$ 
is very different as compared to all the other chains considered 
here. As discussed in the previous section, the Cr isotopes show a 
shape transition between prolate and spherical shapes periodically 
with the neutron number (Fig. 10).  However, the most of the Cr 
isotopes (A=48-76) show only an insignificant difference in the 
deformations of the neutron and proton mean fields, 
although the mean deformation of several Cr isotopes 
rises to about 0.20 or more (see Fig. 10). Only for very 
light Cr isotopes close to the proton drip line and for very heavy 
ones close to the neutron drip line, does $\beta_2 (n) - \beta_2(p)$ 
take a significant value. For $^{38}$Cr (N=14) which has a mean
prolate deformation $\beta_2 \sim 0.29$, the proton deformation
$\beta_2 (p)$ assumes the value $\sim 0.32$ and the neutron
deformation $\beta_2 (n)$ is only 0.23. Thus, the proton mean field is
considerably more deformed than the neutron mean field and drives the
nucleus to a highly deformed shape. Such an effect has not been seen
for proton drip nuclei for other chains. It may, however, be noted 
that although being close to the proton drip line, the nuclei 
$^{38}$Cr and $^{40}$Cr are unbound in the RMF theory.

An interesting but less pronounced difference in the deformations
can be observed for Cr nuclei close to the neutron drip line. 
The nuclei $^{78}$Cr (N=54) and $^{80}$Cr (N=56) are very close 
to the drip line and are prolate deformed. 
The $\beta_2 (n) - \beta_2(p)$ values for these nuclei show that 
the neutron deformation is larger than the corresponding proton 
deformation. This is just contrary to what has been seen for 
Cr nuclei near the proton drip line. In the case of the Cr 
isotopes, the protons and neutrons seem to play complementary roles
at the proton and neutron drip lines, respectively, whereby
deformations at both the drip lines are accentuated by the 
corresponding driping nucleons. Thus, Cr nuclei present an unique 
example where nuclei on the proton drip line as well as on the 
neutron drip line are predicted to be more deformed than those
about the stability line.

\subsection{The Shape-coexistence}

The phenomenon of shape coexistence is known to occur in several
regions of the periodic table. Nuclei in the ground state assume
two different deformations and two minima in the binding energy
occur with a difference of a few hundred keV. Due to a complex
potential-energy landscape in the deformation space, the associated
shapes are usually of oblate and prolate types. The region
of nuclei which we investigate in this work is prone to strong
deformations on both the sides of the spherical shape. That the
isotopes switch the shape from one type of the deformation to
another type on addition of a pair of neutrons is evident from
Figs. 9 and 10. This effect is especially striking for Si and S
chains whereby frequent shape changes are encountered in the
minimum-energy state. Due to this softness of nuclei, it is 
expected that the two different shapes would coexist. 

We show in Table 7 the nuclei which exhibit shape-coexistence
in the RMF theory. The nuclei include an isotope each of Ne and
Ar and several isotopes of Mg, Si and S. The difference in the
binding energy of the prolate minimum and the oblate minimum is shown.
A negative value implies that the prolate minimum lies lower
than the corresponding oblate minimum. For all the Mg isotopes
shown in the table the prolate shape is the lowest energy state 
with the oblate shape lying about 200-400 keV higher except
for $^{42}$Mg where the difference exceeds 1 MeV. Thus,
Mg isotopes $^{26}$Mg, $^{30}$Mg, $^{44}$Mg exhibit the
shape-coexistence truly, wherein the nuclei possess considerable
deformation of both the types.

Several Si isotopes are also shown to provide the occurence of
the shape coexistence. The nuclei $^{26}$Si, $^{30}$Si, $^{38}$Si
and $^{40}$Si are predicted to be prolate deformed in the lowest
energy state and are in coexistence with an oblate shape a few
hundred keV above. Only for $^{32}$Si, the oblate shape is lowest 
in energy. The corresponding deformation for the shape-coexisting
second minimum is very small so as to qualify it for a
spherical shape. Thus, the oblate shape is in coexistence
with the spherical one for $^{32}$Si.

The S isotopes which are predicted to show the shape coexistence
are $^{44}$S, $^{48}$S, $^{50}$S and $^{52}$S. Except for $^{48}$S,
S isotopes are oblate in the lowest energy state with a prolate
shape within 200 keV of the lowest minimum. Here, the case of
$^{44}$S is a celebrated one as this nucleus has the magic
neutron number N=28. The shell effects associated to N=28 are 
predicted to be quenched and the nucleus assumes a deformed shape.
In its lowest energy state, the nucleus is oblate ($\beta_2 = -0.19$).
However, almost at about the same energy (within 30 keV), a highly 
deformed prolate shape ($\beta_2 = 0.38$) coexists in its ground state. 
This is the largest deformation predicted for a sulphur isotope.
It is singular to observe the almost degenerate two different
shapes for this nucleus with the magic neutron number N=28.

\subsection{The Isotope Shifts}
 
We show in Fig. 13 the isotope shifts for nuclei in various isotopic
chains. The isotope shifts have been obtained with respect to a
reference nucleus (shown by a dark point) in each chain as given
by
\begin{equation}
\delta r_c^2 = r_c^2 - r_c^2 (ref)
\end{equation}
It is interesting to note that the general behaviour of the isotope 
shifts as a function of isospin is similar for all the chains
concerned here, i.e. the parabola like response is exhibited
by all the isotopic chains. The minima in the isotope shifts
correspond closely to nuclei near stability line and an upward
trend in going to lighter nuclei implies that the nuclei with
the minimum isotope shifts are smaller in charge radius
even as compared to the lighter neighbours. The deformation
of nuclei contributes partly to an increase in the size of
the lighter isotopes. Such a phenomenon has been observed 
experimentally in general where nuclei are deformed \cite{Ott.89}.
Theoretically, the RMF theory was, for the first time ever, able
to reproduce such a behaviour of the isotopes shifts, for example,
in the isotopic chains of Sr and Kr nuclei \cite{LS.95} and for 
nuclei in the rare-eath region \cite{LS.96}. A similar behaviour 
is predicted here for the light nuclei. Experimental measurements 
on charge radii and isotope shifts are scanty and therefore 
a suitable comparison of our results with data is hindered.

For the highly neutron-deficient nuclei, the isotope shifts rise
considerably above the zero level. A part of the rise to extreme 
left of each curve owes to the nuclei being close to
the proton drip line whereby the charge radius swells due to
a larger proton skin. 

\subsection{Deformation at Drip Lines}

Drip lines are usually encountered near magic numbers. It was 
predicted in the RMF theory that the shell effects associated 
to the magic numbers near the
neutron drip line are strong. This implies that a major gap in energy
should exist for nuclei with particle number above the magic number.
Consequently, nuclei near a major magic number tend to be spherical. 
Hence, it is widely expected that nuclei near a drip line should be
spherical. However, in the present case, nuclei in this region are 
highly deformed and there are indications of quenching of 
the magic number N=28, thus, leading to strong deformations for nuclei
possesing this neutron number. We  have discussed deformation
properties in Figs. 8 and 9 and in Tables 3-5. Here, we summarize
deformed nuclei near drip lines. 

It is interesting to observe from
Figures 8 and 9 that the isotopes at the end of the chains of
Ne, Mg, Si, Ar and Cr (except S and Ti) are deformed. These nuclei
are all close to the neutron drip line and are also in the vicinity
of N=28. Especially, Ne and Mg nuclei near the neutron drip line are
highly prolate deformed ($\beta_2 \sim 0.4$) whereas Si and Ar nuclei near
the neutron drip line are predicted to be highly oblate in shape. For
nuclei above Z=14 (Si) the neutron drip line is encountered at neutron
numbers well above N=28. As discussed in Section 4.6, $^{44}$S with
N=28 is predicted to be highly deformed, whereas S isotopes near
neutron drip (above N=28) show only a moderate deformation. Only for
Ti, isotopes near neutron drip line are spherical as Ti lies in a
traditionally spherical domain of Z=20. However, Cr isotopes near
neutron drip exhibit a reasonably strong deformation. It is noteworthy
that these nuclei occur with a neutron number near N=50 and yet lending
themselves to a well deformed shape. 

Here we have shown that the RMF theory predicts several instances of
strong deformation near the neutron drip line. A comparison of the RMF with
the FRDM results (Figs. 8 and 9) shows that for the case of Ne, Mg
and Cr, both the RMF and FRDM predict strongly prolate shape for
nuclei near the neutron drip line, whereas both predict strongly deformed
oblate shape for Ar and Si nuclei in the vicinity of the neutron drip. 
The significant difference in the two predictions, however, appears
for S isotopes near neutron drip. In the RMF, the S isotopes take up
a spherical shape, whereas the FRDM predicts the corresponding nuclei
to be strongly prolate deformed. The large postive value of $\beta_2$
in the FRDM for $^{54,56}$S arises from an abrupt shape transition
prolate-oblate-prolate at A=52. Such a behaviour is not noticed in 
the RMF predictions.

Examining the proton-rich side of the isotopic chains, we find that 
the RMF theory predicts well-deformed shapes also close to the proton 
drip line. For the Ne and Mg isotopes, the proton drip nuclei go
down to N=8 in the neutron number, thus forcing a spherical mean field.
Thus, the nuclei $^{18}$Ne and $^{20}$Mg are predicted to be
spherical. This is a reminder that N=8 shell strength is maintained
in the sd-shell nuclei. As the proton drip in Ne and Mg nuclei occurs
very close to the stable nuclei, it is difficult to make an assessment
of how the properties of nuclei change near the proton drip line.
We examine other nuclear chains above Z=12 for this purpose. 
For Si, the isotopes $^{22}$Si and $^{24}$Si are near the proton drip 
line. Here, $^{22}$Si is spherical due to N=8, which is already
unbound and is therefore, not a truly drip line nucleus.
However, $^{24}$Si has a proton Fermi energy of $-1.66$ MeV.
It is a proton drip nucleus and is significantly prolate deformed
(see Fig. 8). 

As we go to the S chain, we move away from the influence of N=8 near 
the proton drip line. We obtain $^{28}$S (N=12) as a proton drip nucleus
with $\lambda_p = -1.47$ MeV. It is noteworthy that this nucleus has
a large prolate deformation ($\beta_2 = 0.30$). This serves to
illustrate the point that the mean field near the proton drip line
can be substantially deformed. 

Argon nuclei near the proton drip line show interesting features.
The nuclei $^{30}$Ar and $^{32}$Ar are near the proton drip. 
$^{30}$Ar shows a well-deformed prolate shape ($\beta_2 \sim 0.23$) 
in its ground state. However, at the same time there co-exists 
an oblate deformed ($\beta_2 \sim -0.14$) minimum about 0.5 MeV above
the prolate state. Thus, the shape coexistence seems to be preserved
also near a drip line. $^{32}$Ar is also a proton drip
nucleus, having an oblate deformation $\beta_2 \sim -0.16$.
In contrast, Ti and Cr nuclei near proton drip are overwhelmingly 
spherical due to a strong influence from the magic number Z=20.

\section{Summary and Conclusions}

We have made an exhaustive study of the ground-state properties of
nuclei in the light mass region. The study has covered complete
isotopic chains encompassing nuclei near the stability line 
as well as those close to the neutron and proton drip lines.
As nuclei in this region of the periodic table have significant
deformation, we have employed an axially deformed basis to investigate
the structure of nuclei in the RMF theory. We have used the force
NL-SH which has been shown to describe the properties of nuclei all
along the periodic table. The asymmetry energy plays an important role
for nuclei far away from the stability line. In this respect, the
force NL-SH which has a proper asymmetry energy has been 
shown to perform very well for nuclei with large isospins. 

The ground-state binding energies and charge radii obtained in 
the RMF theory show a good agreement with the data where 
available. The neutron radii for very neutron-rich isotopes are 
found to exhibit an unusual increase as compared to their 
lighter counterparts. The increase in the neutron $rms$ radius
is more apparent above a major closed shell, thus reflecting the
importance of the shell effects. Nuclei near the neutron drip line 
show signs of a neutron halo as the size of a nucleus swells
significantly due to the last neutron orbitals being close to the
continuum. However, the corresponding proton $rms$ radius at the
proton drip line does not increase as much due to the reason that 
the Coulomb repulsion brings about the proton drip with only a 
marginal difference in the neutron and proton number. Consequently,
the proton halo in nuclei near the proton drip line is suppressed
considerably as compared to a larger neutron halo expected near 
neutron drip line.

It is shown that nuclei in this region are strongly deformed.
This deformation persists also for nuclei away from the stability 
line. Many nuclei near the drip lines are predicted to be 
strongly deformed. It is also predicted that the phenomenon of shape 
coexistence occurs in several nuclei in these isotopic chains.
This is due to a complex evolution of the potential energy landscape
in the deformation degrees of freedom of nuclei as a function of 
nuclear isospin. Such a feature is also exhibited by nuclei near 
the neutron drip line. Consequently, several isotopes 
of Mg, Si and S near the neutron drip line are shown to coexist in a 
well-deformed prolate and an oblate shape. The two states differ
by a few keV in energy. In this respect, the nucleus $^{44}$S has 
shown a most remarkable character that two minimum-energy states
of this nucleus, one a highly prolate and another well-deformed
oblate, are almost degenerate in energy. 

Examining the effect of the shell closures on deformation,
it is shown that major magic numbers such as 8, 20 and 50 respect 
the shell effects and lead to nuclei which are spherical. This effect 
is observable for nuclei about the stability line as well as near the 
drip lines. This shows that the major shell closures keep their magic
nature even in going to the drip lines. In a broader sense, the
shell effects near the drip lines are maintained as strong.
This is in accord with our earlier assertion that the shell effects
near the neutron drip line in medium-heavy nuclei remain strong 
\cite{SLH.94}. 
 
The magic number N=28, however, exhibits different properties. 
For the isotopic chains below Ca, it is shown that the neutron number
N=28 loses its magic character and that the associated shell effects
are washed out, implying that the shell gap at N=28 is reduced  
significantly. Consequently, the corresponding nuclei deform
strongly. The case of $^{44}$S is shown to be especially interesting 
as this nucleus with N=28 assumes a strongly prolate and a strongly 
oblate shape which coexist in the ground state, as mentioned above.

We have employed the BCS pairing in the present study. For nuclei
close to and not too far from the stability line, the BCS approach
provides a reasonably good description of the pairing properties. 
However, in going to nuclei near the drip lines, coupling to continuum
needs to be taken into account. Therefore, for such nuclei, the
BCS pairing has limitations in that respect. The most appropriate 
framework for an improved prediction of such aspects is the 
relativistic Hartree-Bogolieubov approach in coordinate space. 
An application of the RHB approach for deformed nuclei has still not
been accomplished and is work for future.

\section{Acknowledgment}

One of the authors (GAL) acknowledges support by the European
Union under contract No. TMR-EU/ERB/FMBCICT-950216, by the
Bundesministerium fuer Bildung und Forschung under the project
06 TM 743 (6) and by the Greek Secretariat of Research and
Technology under contract PENED/1981.

\newpage
\leftline{\Large {\bf Figure Captions}}
\vskip 1 true cm
\begin{description}
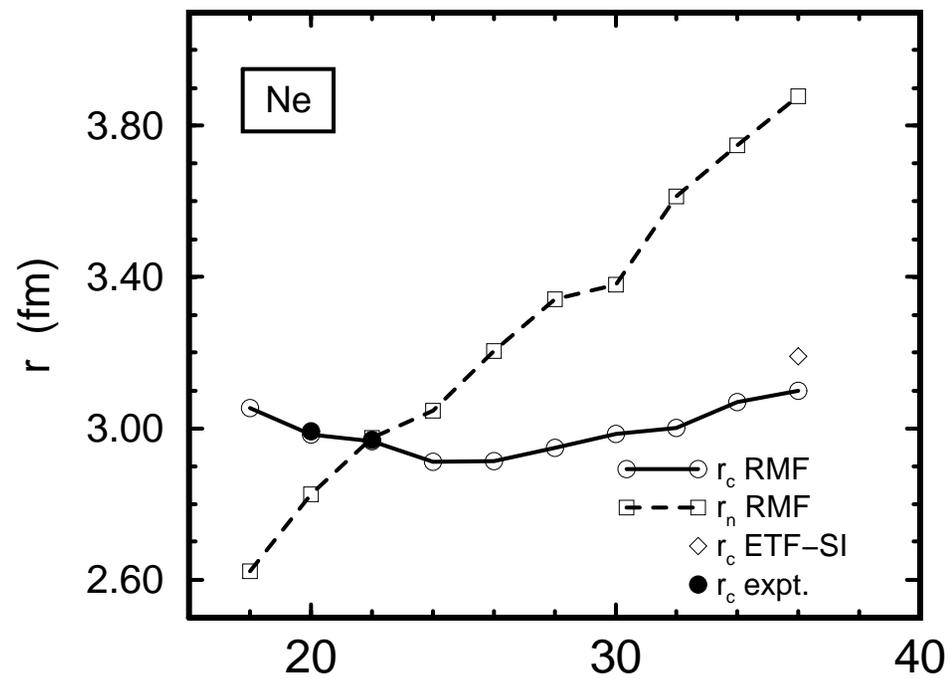
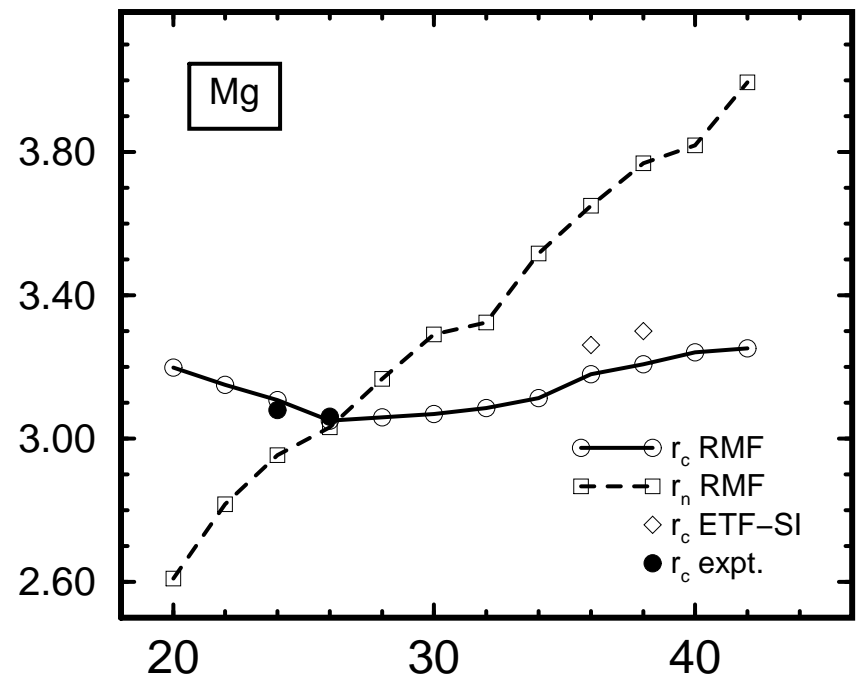
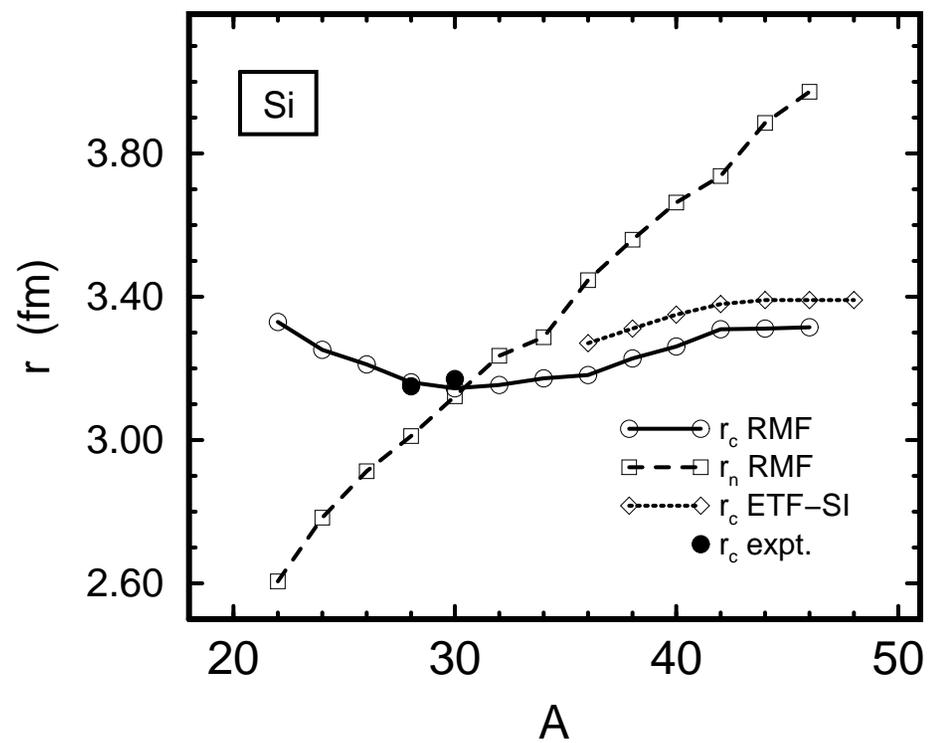
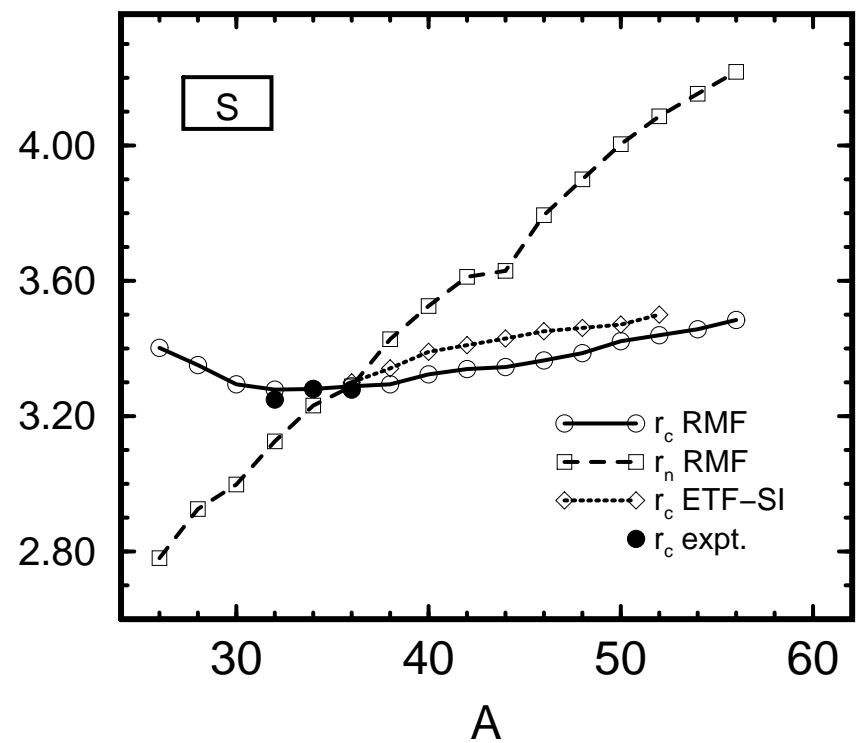
\item[Fig. 1] The charge ($r_c$) and neutron ($r_n$) $rms$ radii
of nuclei obtained in the deformed RMF calculations using the force 
NL-SH. The charge radius has been obtained by folding the 
finite proton size on to the proton $rms$ radius. A few experimental 
charge radii available and some results  from the predictions of 
ETF-SI as available are also shown for comparison.

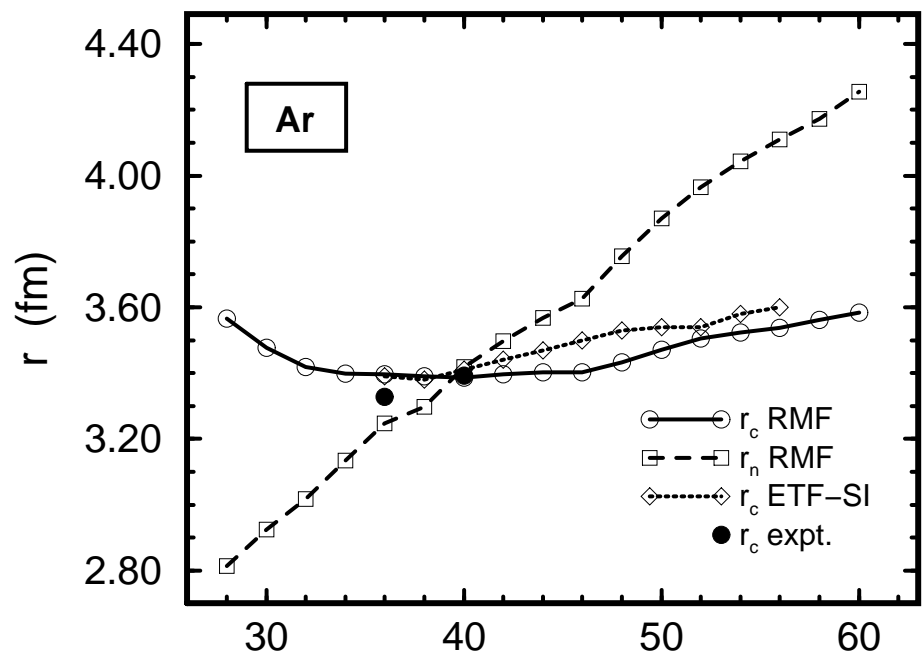
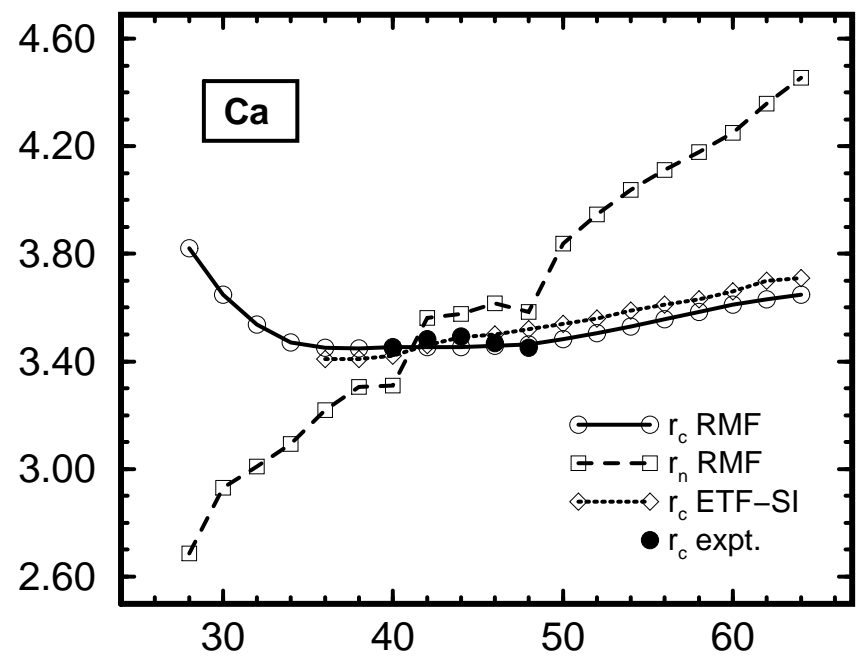
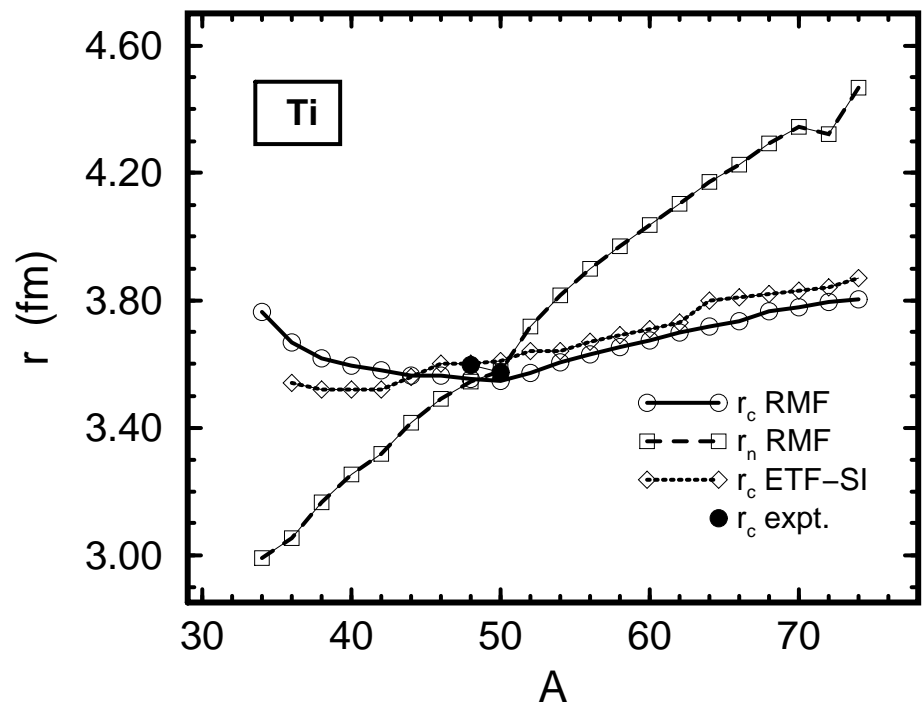
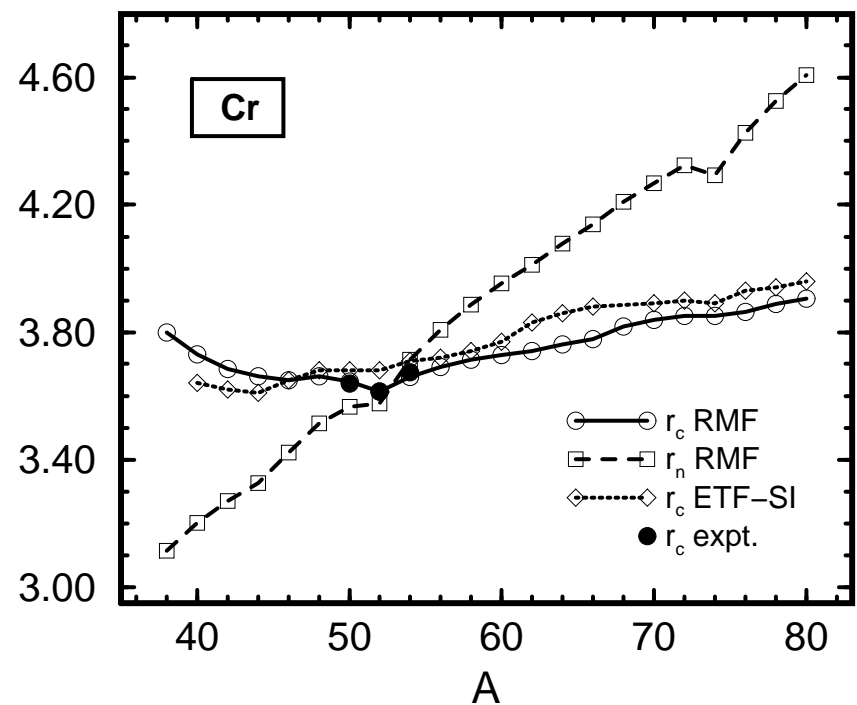
\item[Fig. 2] The same as in Fig. 1, for other isotopic chains.

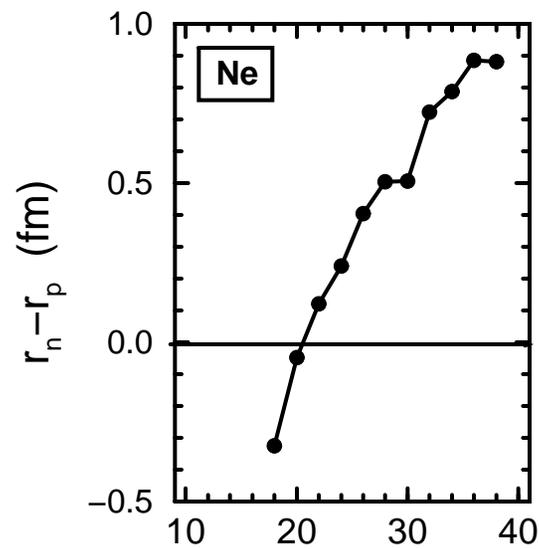
\item[Fig. 3] The neutron skin thickness (proton skin thickness, 
respectively) ($r_n - r_p$) for neutron-rich and proton-rich nuclei
in various isotopic chains. The effect of the Coulomb force in 
expanding the proton $rms$ radius rapidly can be visualized in a 
stronger slope of the curve for proton-rich nuclei.

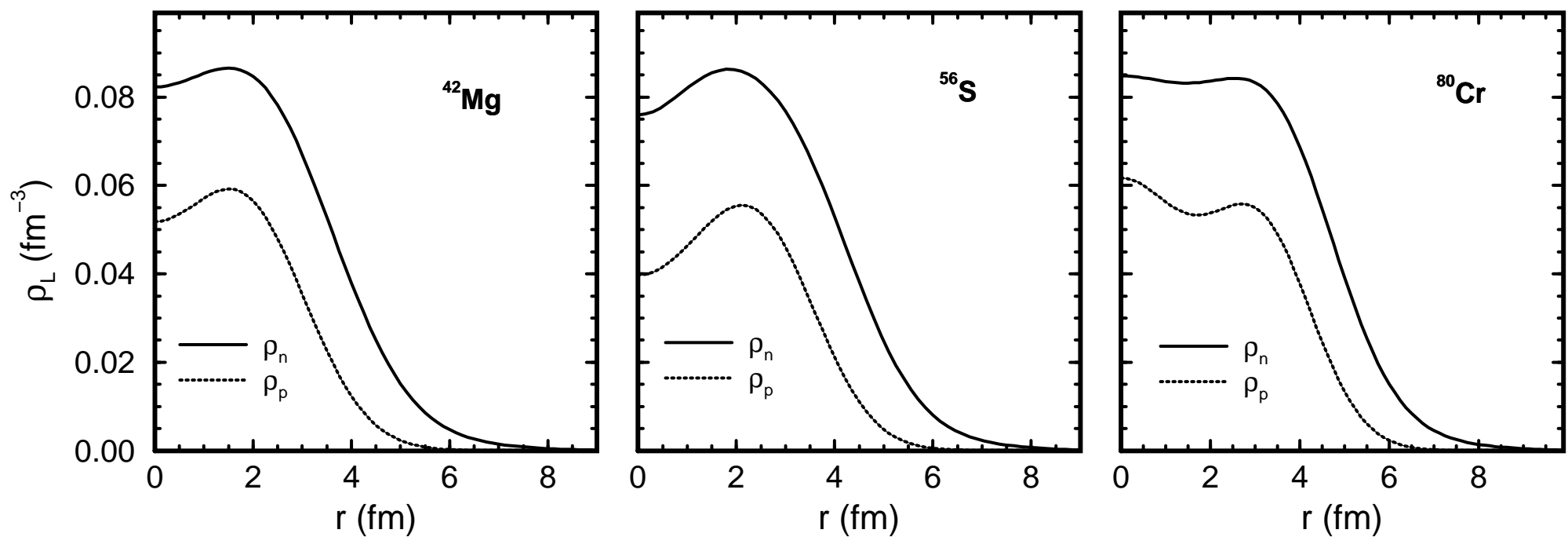
\item[Fig. 4] The L=0 component of the neutron and proton vector
(baryonic) densities for a few nuclei near the neutron drip line.
A staggeringly large neutron halo and a higher neutron density
in the interior as compared to protons is present.

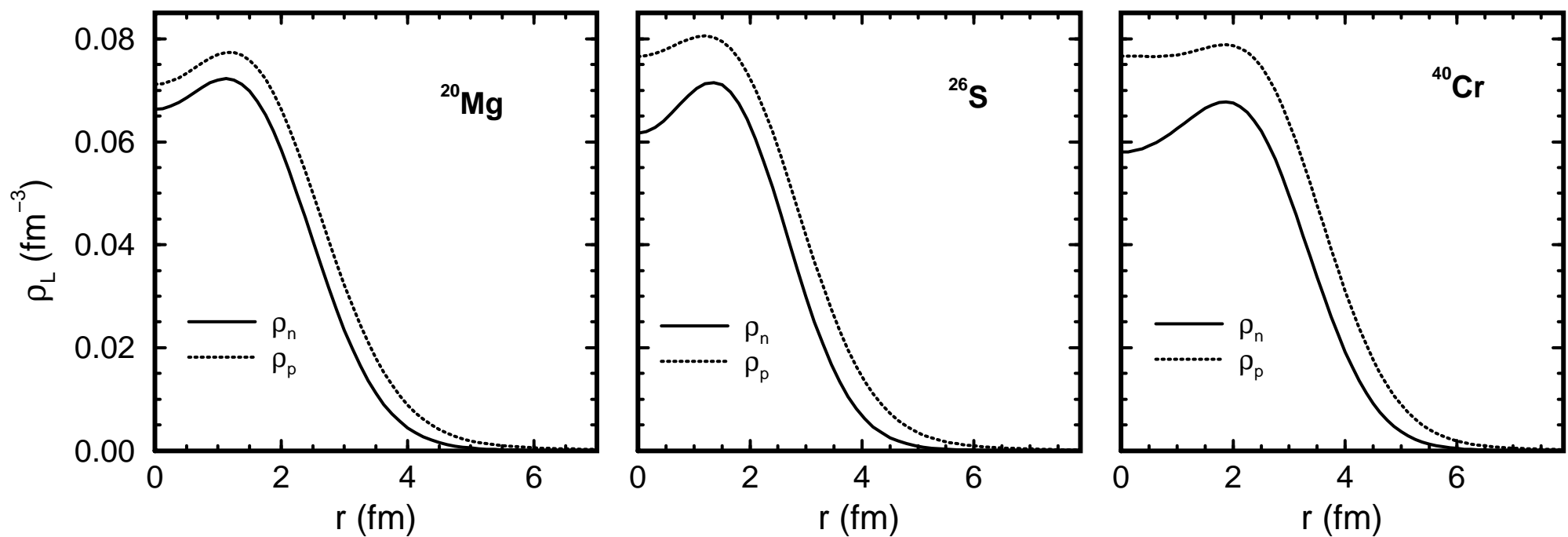
\item[Fig. 5] The same as in Fig. 4 for nuclei near proton drip.
The proton skin thickness and the formation of a proton halo
is subdued due to the Coulomb barrier.

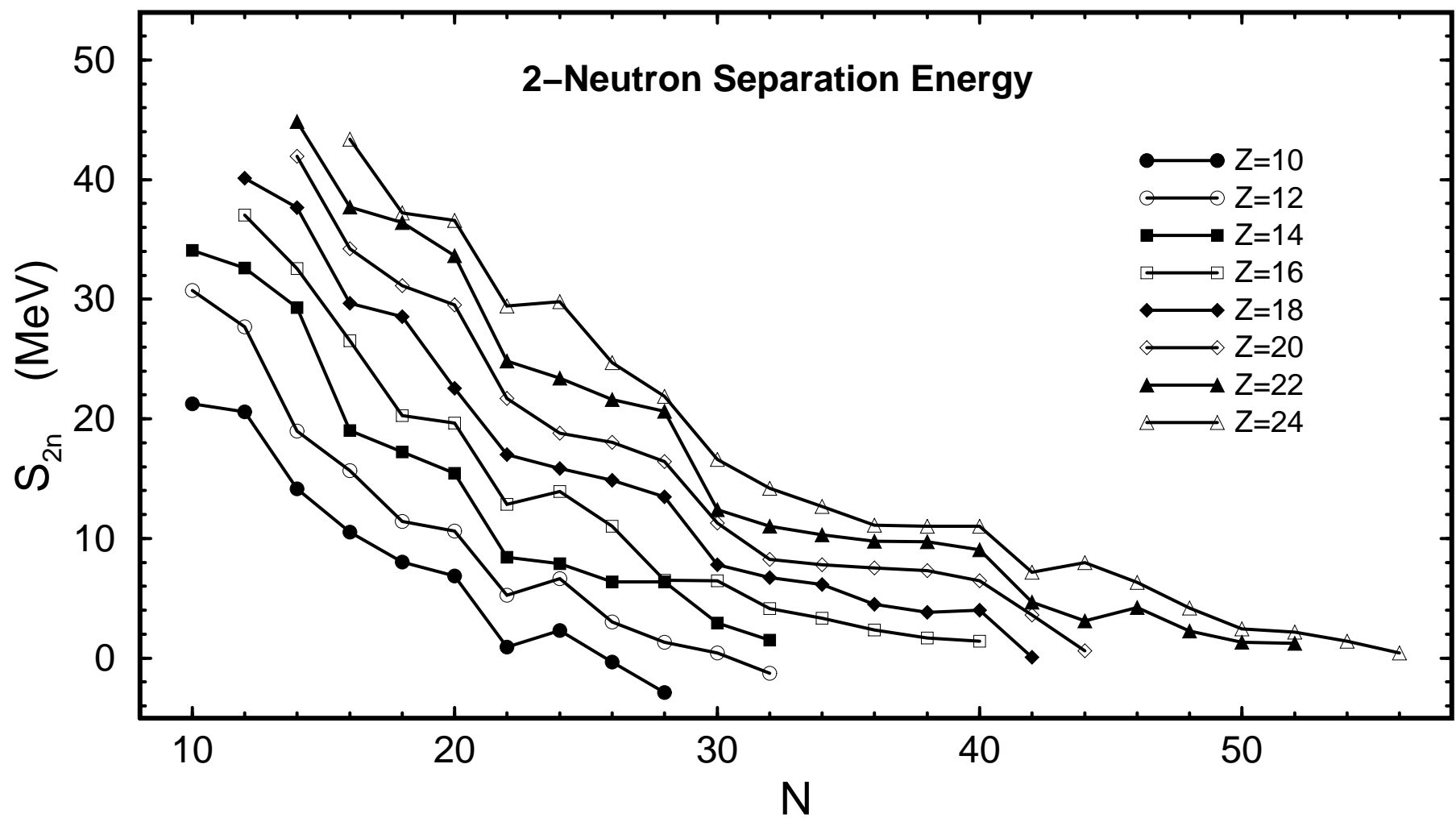
\item[Fig. 6] The 2-neutron separation energy in the RMF theory. 

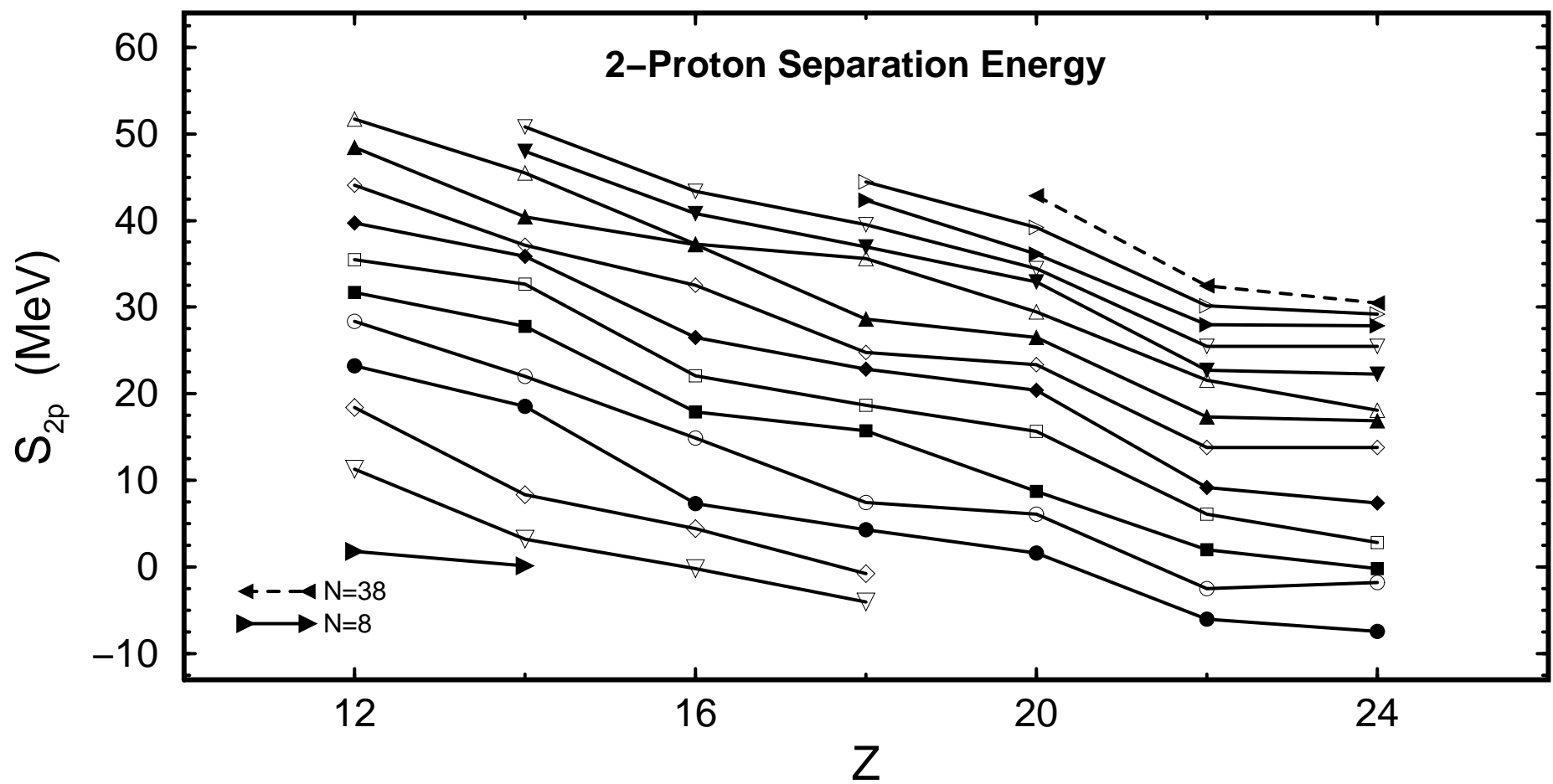
\item[Fig. 7] The 2-proton separation energy in the RMF theory.

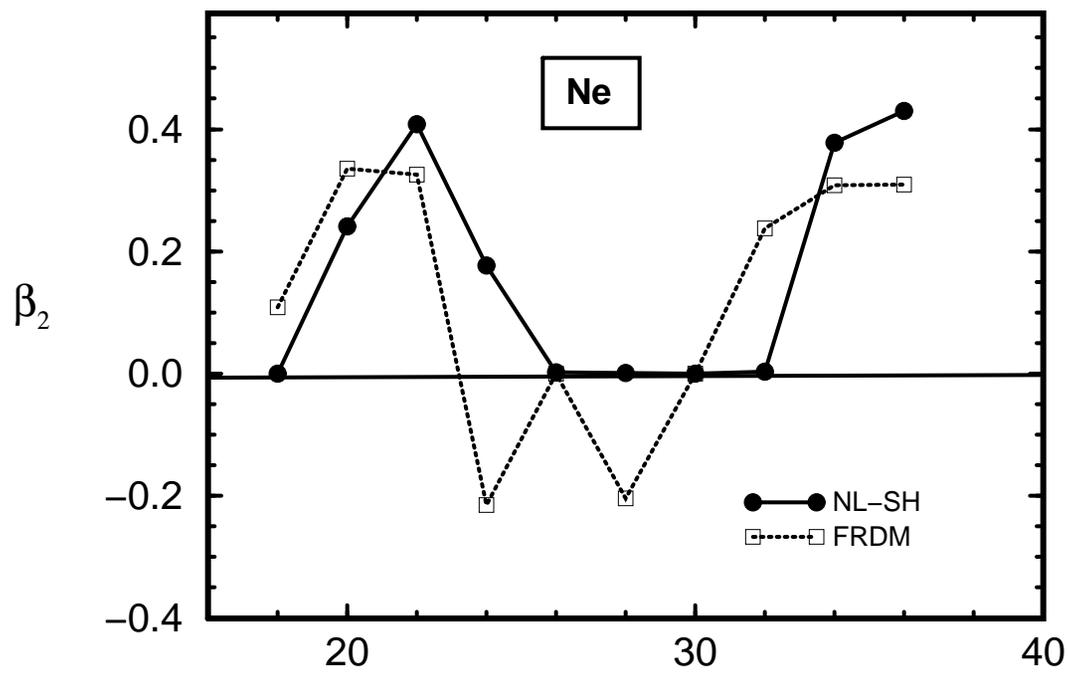
\item[Fig. 8] The quadrupole deformation ($\beta_2$) in the RMF
theory. A comparion is made with the results from FRDM and ETF-SI
where available. 

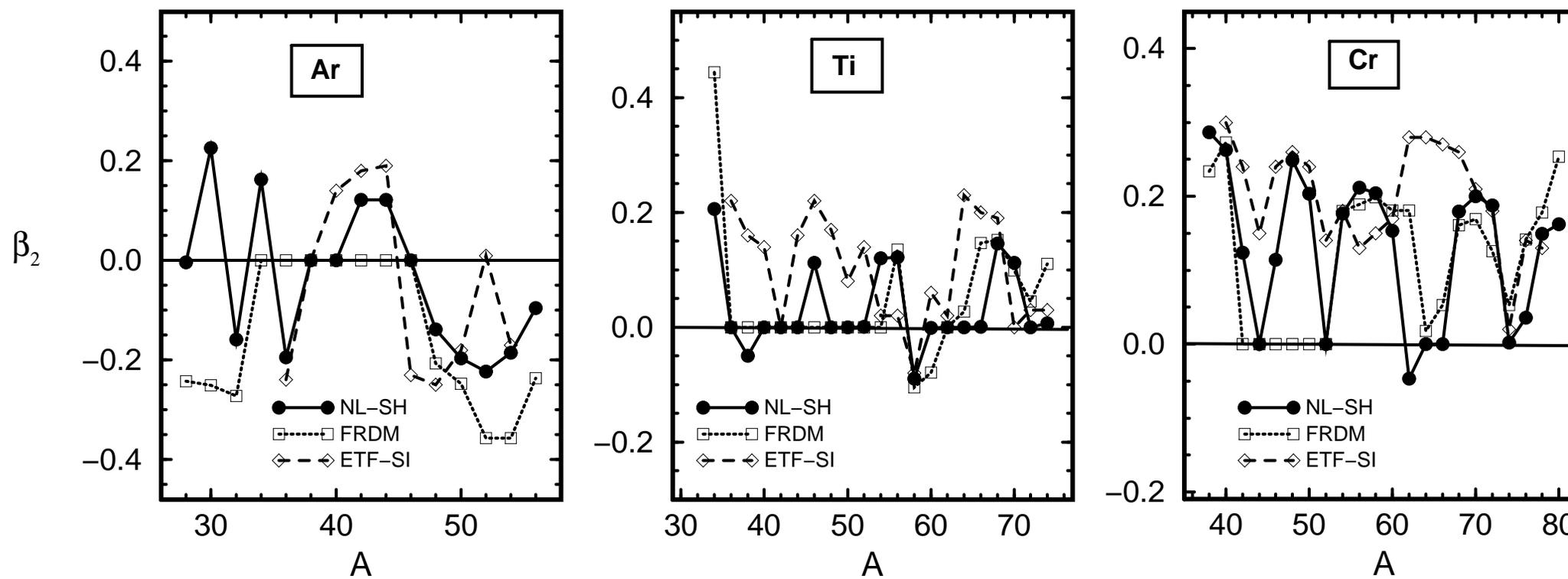
\item[Fig. 9] The same as in Fig. 8 for other isotopic chains.

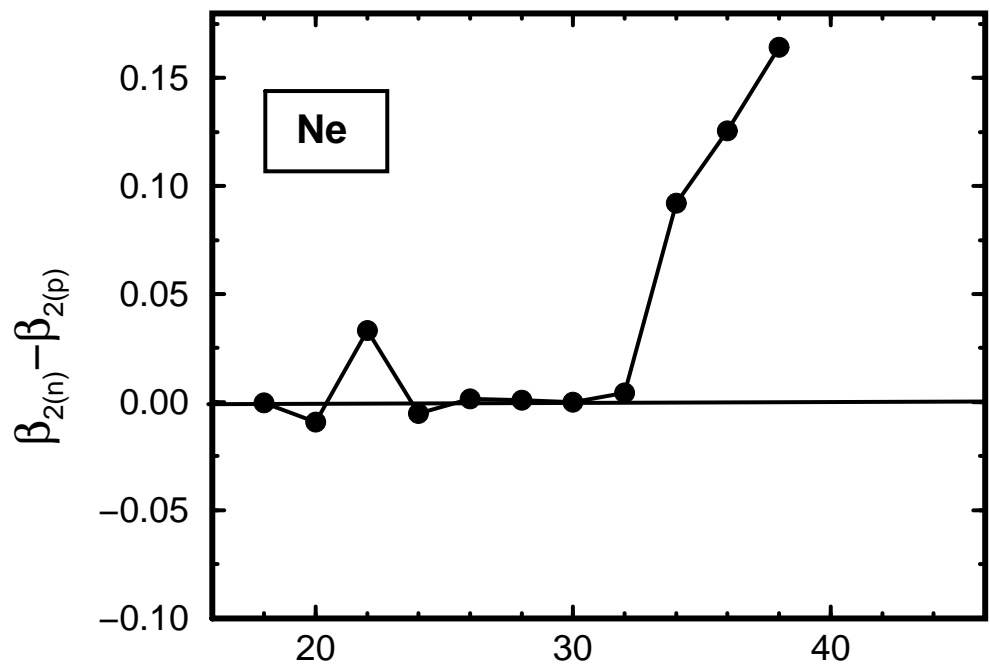
\item[Fig. 10] The difference in the quadrupole deformations of
the neutron and proton mean-fields in the RMF theory.

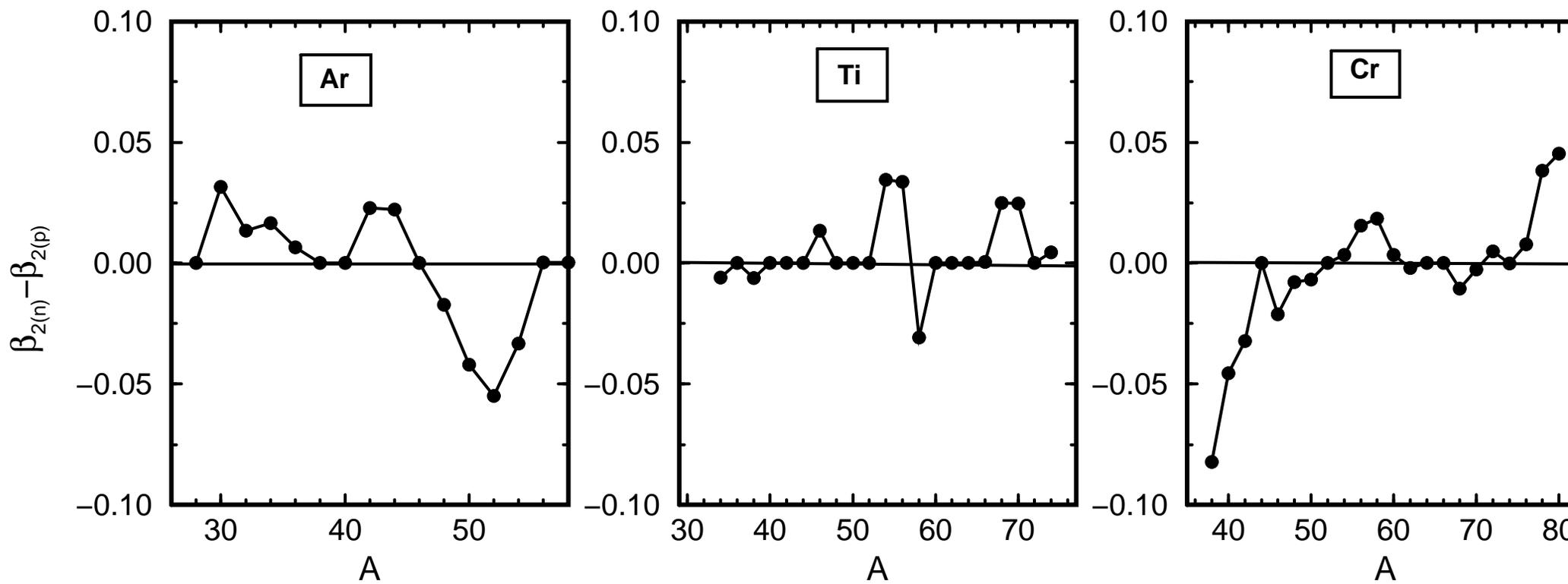
\item[Fig. 11] The same as in Fig. 10 for other isotopic chains.

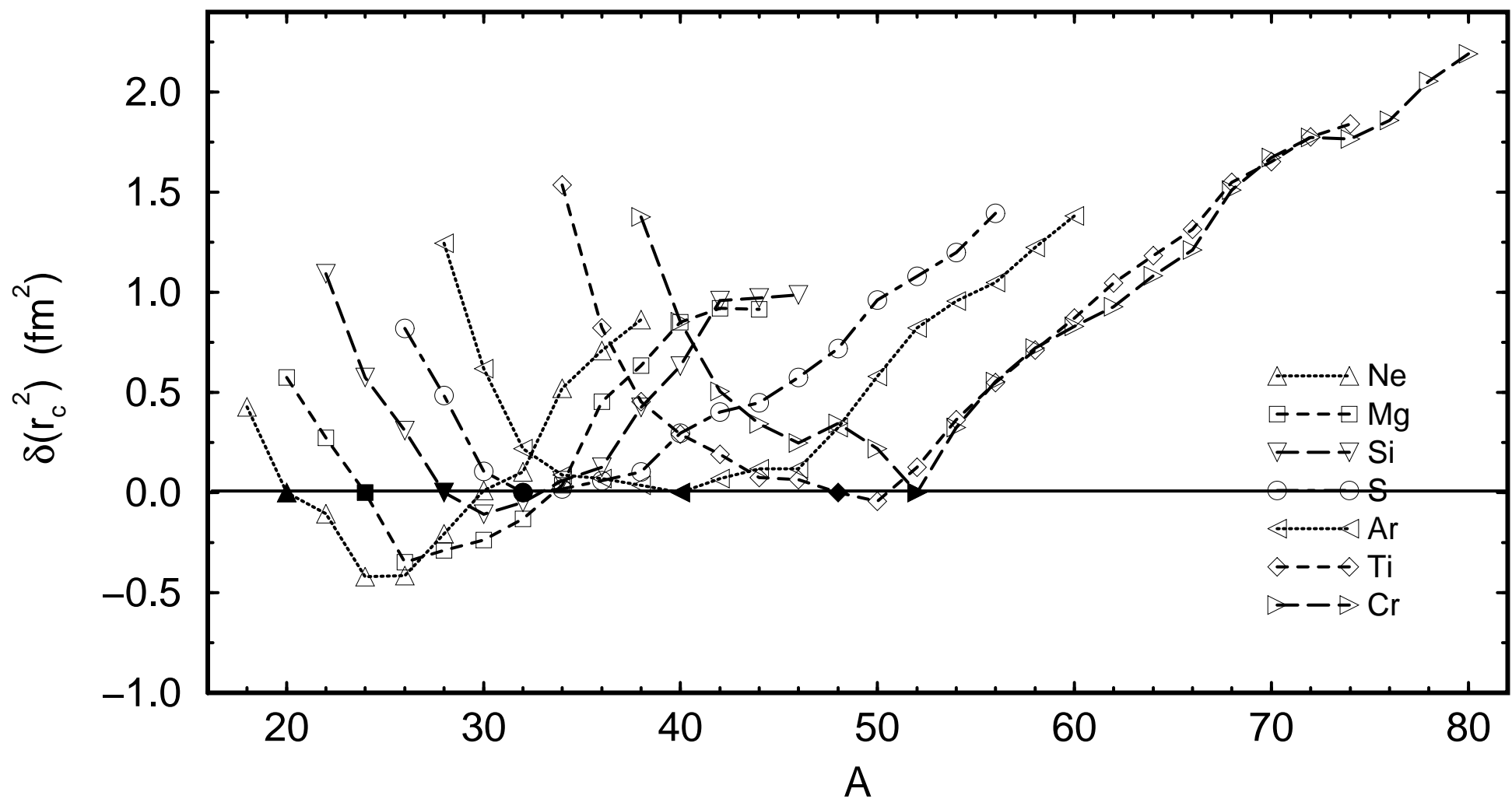
\item[Fig. 12] The isotope shifts obtained in the RMF theory
for various isotopic chains. The reference nucleus chosen for each
chain for calculating the difference is shown by dark points.

\end{description}

\begin{table}
\vspace{-0.5cm}
\begin{center}
\caption{\sf The binding energies (in MeV) for even-even Ne and Mg
Si and S  isotopes obtained with the force NL-SH. The predictions from 
the mass models FRDM and ETF-SI (wherever available) are also shown 
for comparison. The empirical values (expt.) available are also shown.}
\begin{tabular}{l c c c c c c c c c c  l}
\hline
& &     & Ne Nuclei&  &      & &    & Mg Nuclei   &    &\\  
&A& RMF & FRDM& ETF-SI&expt. &A& RMF& FRDM& ETF-SI&expt&\\
\hline
&18&134.10&134.25&-&132.14   &20&135.90&135.97&-&134.35&\\
&20&155.38&161.32&-&160.64   &22&166.65&169.68&-&168.57&\\
&22&175.96&178.44&-&177.77   &24&194.34&198.40&-&198.26& \\
&24&190.12&191.87&-&191.84   &26&213.34&216.86&-&216.68& \\
&26&200.66&201.01&-&201.60   &28&229.01&231.11&-&231.63& \\ 
&28&208.71&208.15&-&206.89   &30&240.43&241.89&-&241.63& \\ 
&30&215.57&211.88&-&212.08   &32&251.03&250.32&-&249.69& \\
&32&216.47&217.14&-&213.28   &34&256.28&257.15&-&256.59& \\
&34&218.80&217.04&-& -       &36&262.90&262.02&259.38&260.27& \\ 
&36&218.47&213.97&207.35&-   &38&265.93&264.58&260.92&-& \\
&38&215.59&214.04&-&-        &40&267.28&268.05&-&-& \\
&  &      &      & &         &42&267.71&264.09&-&-& \\
&  &      &      & &         &44&266.45&267.69&-&-& \\
\hline\hline
& &     & Si Nuclei&  &      & &    & S Nuclei   &    &\\ 
&A& RMF & FRDM& ETF-SI&expt. &A& RMF& FRDM& ETF-SI&expt&\\
\hline
&22&136.00&133.45&-&134.45   &26&169.69&169.60&-&171.37&\\
&24&170.19&172.94&-&172.00   &28&206.74&209.60&-&209.41&\\
&26&202.70&207.25&-&206.05   &30&239.32&241.92&-&243.69&\\
&28&231.99&236.15&-&236.54   &32&265.84&269.52&-&271.78&\\
&30&251.00&253.67&-&255.62   &34&286.12&290.23&-&291.84&\\
&32&268.24&269.63&-&271.41   &36&305.75&308.00&308.15&308.71&\\
&34&283.69&282.98&-&283.43   &38&318.61&321.97&321.19&321.05&\\
&36&292.13&292.99&291.82&292.02 &40&332.51&334.22&333.14&333.18&\\
&38&300.16&301.49&299.76&299.50 &42&343.55&344.27&343.08&343.72&\\
&40&306.36&307.08&305.80&306.50 &44&350.03&352.00&351.52&353.50&\\
&42&312.74&315.17&311.54&-   &46&356.51&357.05&357.27&-&\\   
&44&315.68&315.02&313.59&-   &48&360.63&360.86&361.61&-&\\
&46&317.28&318.05&315.03&-   &50&363.97&363.40&363.55&-&\\
&  &      &      &      &    &52&366.32&362.71&360.89&-&\\
&  &      &      &      &    &54&368.02&362.60&-&-&\\
&  &      &      &      &    &56&369.60&361.86&-&-&\\
\hline
\end{tabular}
\end{center}
\end{table} 
\vfill
\begin{table}
\vspace{-2.2cm}
\begin{center}
\caption{\sf The binding energies of Ti, Cr and Ar isotopes. Refer to
the caption of Table 1 for details.}
\begin{tabular}{l c c c c c c c c c c l}
\hline
& &     & Ti Nuclei&  &   &   &  &  Cr Nuclei       &  &\\ 
&A& RMF &FRDM&ETF-SI&expt.&A&RMF&FRDM&ETF-SI&expt&\\
\hline
&34&194.33&193.77&-&-&38&231.72&232.29&-&-&\\
&36&239.18&239.83&-&240.66&40&275.10&275.68&277.58&-&\\
&38&276.90&280.29&280.80&280.40&42&312.33&314.78&317.99&314.23&\\
&40&312.52&316.04&315.84&314.49&44&348.94&351.02&351.26&349.90&\\
&42&346.15&348.69&347.17&346.91&46&378.63&382.63&381.81&381.98&\\
&44&370.97&376.76&373.23&375.47&48&408.15&411.16&409.65&411.46&\\
&46&394.39&399.17&397.07&398.19&50&432.84&435.39&433.99&435.04&\\
&48&416.00&420.10&418.21&418.70&52&454.74&457.29&455.64&456.35&\\
&50&436.64&438.66&437.26&437.78&54&471.34&474.55&474.08&474.00&\\
&52&449.06&452.72&452.40&451.96&56&485.55&489.02&489.32&488.51&\\
&54&460.09&464.17&465.24&464.25&58&498.23&501.84&501.86&501.25&\\
&56&470.40&474.37&474.89&473.91&60&509.35&513.42&512.51&512.33&\\
&58&480.17&483.05&482.33&-&62&520.38&522.83&522.15&-&\\  
&60&489.90&491.02&488.77&-&64&531.42&532.74&530.79&-&\\
&62&498.96&497.89&494.21&-&66&538.61&540.26&538.13&-&\\
&64&503.63&503.43&498.86&-&68&546.82&547.42&544.38&-&\\
&66&506.73&507.75&502.70&-&70&553.15&553.58&549.82&-&\\
&68&510.98&511.23&505.74&-&72&557.32&558.23&554.06&-&\\
&70&513.26&514.13&507.78&-&74&559.78&562.49&557.31&-&\\
&72&514.59&516.46&509.03&-&76&561.94&563.44&558.85&-&\\
&74&515.92&515.64&508.97&-&78&563.34&563.50&559.69&-&\\
&  &      &      &      &      &80&563.75&561.79& &-&\\
\hline\hline
& &     &           &    & Ar Nuclei   &      &     &        &     & \\ 
& &     &A&RMF & FRDM &ETF-SI&expt. &     &        &     \\
& &     &28&165.64&166.11&-     &-     &   &  &   \\ 
& &     &30&205.96&208.11&-     &207.97&   &  &    \\
& &     &32&243.61&246.32&-     &246.38&   &  &    \\
& &     &34&273.29&277.37&-     &278.72&   &  &    \\
& &     &36&302.83&306.05&304.19&306.72&   &  &    \\
& &     &38&324.39&328.01&326.23&327.34&   &  &    \\
& &     &40&341.42&345.30&343.77&343.81&   &  &    \\
& &     &42&357.27&360.53&359.62&359.34&   &  &    \\
& &     &44&372.15&374.36&373.46&373.32&   &  &    \\
& &     &46&385.62&386.19&386.00&386.92&   &  &    \\
& &     &48&393.44&394.58&394.94&396.56&   &  &    \\
& &     &50&400.16&400.96&401.99&-     &   &  &    \\
& &     &52&406.33&405.92&406.33&-     &   &  &    \\
& &     &54&410.81&408.91&407.07&-     &   &  &    \\
& &     &56&414.70&412.93&407.21&-     &   &  &    \\
& &     &58&418.60&413.57&-     &-     &   &  &    \\
& &     &60&418.80&414.64&-     &-     &   &  &    \\
\hline
\end{tabular}
\end{center}
\end{table}
\noindent\begin{table}
\vspace{-0.5cm}
\begin{center}
\caption{\sf The $\beta_{2}$ and $\beta_{4}$ deformation
parameters calculated in the RMF theory with the NL-SH
force for various even-even Ne (Z=10) Mg (Z=12) and Si (Z=14) isotopes. 
Predictions from the mass models FRDM and ETF-SI are also shown for 
comparison. For the values given in the parenthesis, refer to the text.}
\begin{tabular}{ll l l l c c c c c c  l}
\hline
&   &   &    & $\beta_{2}$ &  & &  &  $\beta_{4}$ &    &\\
\hline
& Z & A & N & RMF & FRDM & ETF-SI & & RMF& FRDM& ETF-SI&\\
\hline
&10&18&~8& 0.002         & 0.109 &- & &0.000 & 0.150& -&\\
&  &20&10& 0.241         & 0.335 &- & &0.059 & 0.428& -&\\  
&  &22&12& 0.408         & 0.326 &- & &0.065 & 0.225& -&\\
&  &24&14& 0.178 (-0.107)&-0.215 &- & &0.021 & 0.155& -&\\ 
&  &26&16& 0.002 & 0.000 & 0.370 & &0.000 &-0.014& 0.000 &\\ 
&  &28&18& 0.001 &-0.204 &- & &0.000 &-0.127& -&\\  
&  &30&20& 0.000 & 0.000 &- & &0.000 &-0.014& -&\\ 
&  &32&22& 0.004 & 0.238 &- & &0.000 & 0.281& -&\\
&  &34&24& 0.377 & 0.308 &- & &0.194 & 0.220& -&\\ 
&  &36&26& 0.430 & 0.309 &- & &0.079 & 0.065& -&\\ 
&  &38&28& 0.443 &-0.292 &- & &-0.088 & 0.196& -&\\   
\hline
&12&20&~8& 0.000 & 0.147 &- & &0.001&-0.091&-&\\
&  &22&10& 0.426 & 0.326 &- & &0.080&0.225&-&\\ 
&  &24&12& 0.465 (-0.204) & 0.374 &- & &0.005&-0.053&-&\\   
&  &26&14& 0.298 (-0.219) &-0.310 &- & &0.000&0.186&-&\\ 
&  &28&16& 0.288 (-0.164) & 0.323 &- & &-0.027& -0.136&-&\\
&  &30&18& 0.189 (-0.134) &-0.222 &- & &-0.017& -0.112&-&\\ 
&  &32&20& 0.000 & 0.000 &- & & 0.000& 0.000& -&\\   
&  &34&22& 0.173 & 0.406 &- & & 0.042& 0.062& -&\\
&  &36&24& 0.383 (-0.165) & 0.328 &0.350 & & 0.071& 0.085& 0.070&\\
&  &38&26& 0.413 (-0.246) & 0.314 &0.390 & & 0.020& -0.037&-0.010&\\ 
&  &40&28& 0.449 (-0.354) &-0.290 &- & &-0.086& 0.211& -&\\   
&  &42&30& 0.414 (-0.300) & 0.265 &- & &-0.056& -0.015&-&\\  
&  &44&32& 0.317 (-0.243) & 0.175 &- & &-0.049& -0.317&-&\\  
\hline
&14&22&~8&  0.000&  0.000&-&  &0.000&0.000&-&\\ 
&  &24&10&  0.240 (-0.140) & -0.242&- & &0.037&0.202&-&\\
&  &26&12&  0.330 (-0.230) & -0.353&- & &-0.002&0.226&-&\\
&  &28&14& -0.287& -0.478&- & &0.070&0.250&-&\\ 
&  &30&16& -0.185 (0.122) &  0.000&- & &0.007&0.000&-&\\
&  &32&18&  0.028 (-0.151) &  0.000&- & &0.004&0.000&-&\\
&  &34&20&  0.000&  0.000&- & &0.000&0.000&-&\\
&  &36&22&  0.000&  0.000&0.180 & &0.000&0.000&0.050&\\ 
&  &38&24&  0.248 (-0.155) &  0.234&0.260 & &0.048&0.094&0.050&\\ 
&  &40&26&  0.291 (-0.240) & -0.592&-0.310 & &0.002&0.143&0.040&\\ 
&  &42&28& -0.342 (0.425) & -0.321&-0.350 & &0.164&0.218&0.050&\\  
&  &44&30& -0.304 (0.137) & -0.263&-0.300 & &0.083&0.071&-0.030&\\ 
&  &46&32& -0.255&  0.023&-0.180 & &0.013&-0.291&-0.070&\\
\hline
\end{tabular}
\end{center}
\end{table}
\noindent\begin{table}
\vspace{-1.cm}
\begin{center}
\caption{\sf As in Table but for S (Z=16) and Ar (Z=18) isotopes.} 
\begin{tabular}{ll l l l  c c c c c c  l}
\hline
&   &   &    & $\beta_{2}$ &  & &  &  $\beta_{4}$ &    &\\
\hline
& Z&A & N & RMF & FRDM & ETF-SI & & RMF& FRDM& ETF-SI&\\
\hline
&16&26&10& 0.004 & 0.000 &-      & &0.000 & 0.000& -    &\\
&  &28&12& 0.307 & 0.346 &-      & &-0.011&-0.159& -    &\\ 
&  &30&14&-0.200 & 0.000 &-      & &0.016 & 0.000& -    &\\
&  &32&16& 0.235 & 0.000 &-      & &-0.042& 0.000& -    &\\  
&  &34&18& 0.146 & 0.000 &-      & &-0.023& 0.000& -    &\\
&  &36&20& 0.000 & 0.000 & 0.020 & &0.000 & 0.000&-0.030&\\
&  &38&22&-0.043 & 0.000 & 0.160 & &0.001 & 0.000& 0.010&\\ 
&  &40&24& 0.242 & 0.254 & 0.260 & &0.008 &-0.001& 0.020&\\
&  &42&26& 0.259 & 0.249 & 0.250 & &-0.027 &-0.076&-0.020&\\ 
&  &44&28&-0.195 (0.375) & 0.000 &-0.260 & &0.047  &-0.008& 0.030&\\
&  &46&30& 0.209 & 0.219 &-0.250 & &-0.064 &-0.048&-0.050&\\
&  &48&32& 0.201 (-0.203) & 0.241 &-0.180 & &-0.029 &-0.086&-0.070&\\ 
&  &50&34&-0.218 (0.169) & 0.255 & 0.010 & &-0.038 &-0.139& 0.000&\\
&  &52&36&-0.168 (0.130) &-0.406 &-0.130 & &-0.038 &-0.035& 0.030&\\
&  &54&38& 0.078 (-0.087)  & 0.434 & -     & &-0.019 &-0.051&  -   &\\
&  &56&40& 0.000 & 0.461 & -     & & 0.000 &-0.102&  -   &\\ 
\hline
&18&28&10& 0.000 &-0.243 &     & & 0.000  &-0.244&-    &\\
&  &30&12& 0.226 (-0.138) &-0.251 &-     & &-0.019  &-0.168&-    &\\  
&  &32&14&-0.159 &-0.272 &     & &-0.013  &-0.114&-    &\\
&  &34&16& 0.163 & 0.000 &     & &-0.025  & 0.000&-    &\\ 
&  &36&18&-0.195 & 0.000 &-0.240& &-0.057  & 0.000&-0.060&\\ 
&  &38&20& 0.000 & 0.000 & 0.000& & 0.000  & 0.000&-0.010&\\  
&  &40&22& 0.000 & 0.000 & 0.140& & 0.000  & 0.000& 0.000&\\
&  &42&24& 0.122 & 0.000 & 0.180& & 0.001  & 0.000& 0.010&\\
&  &44&26& 0.122 & 0.000 & 0.190& & 0.010  & 0.000& 0.000&\\ 
&  &46&28& 0.000 & 0.000 &-0.230& & 0.000  & 0.000& 0.030&\\ 
&  &48&30&-0.138 &-0.207 &-0.250& &-0.009  &-0.067&-0.050&\\
&  &50&32&-0.196 (0.088) &-0.248 &-0.180& &-0.024  &-0.089&-0.070&\\
&  &52&34&-0.224 (0.095) &-0.306 & 0.010& &-0.039  &-0.105& 0.000&\\
&  &54&36&-0.186 (0.066) &-0.357 &-0.170& &-0.051  &-0.056&-0.040&\\
&  &56&38&-0.100 &-0.237 &-0.160& &-0.027  &-0.126&-0.050&\\ 
&  &58&40& 0.001 &-0.255 &  -   & & 0.000  &-0.123& -& \\
&  &60&42& 0.000 &-0.285 &  -   & & 0.000  &-0.161& -& \\ 
\hline
\end{tabular}
\end{center}
\end{table}
\vfill
\noindent\begin{table}
\vspace{-2.5cm}
\begin{center}
\caption{\sf As in Table but for Ti (Z=22) and Cr (Z=24) isotopes.}
\begin{tabular}{ll c c c c c c c c c  l}
\hline\
&   &   &    & $\beta_{2}$ &  & &  &  $\beta_{4}$ &    &\\
\hline
& Z &A& N & RMF & FRDM & ETF-SI & & RMF& FRDM& ETF-SI&\\
\hline
&22&34&12& 0.206 & 0.444 & -    & &0.051& 0.074& -   &\\
&  &36&14& 0.000 & 0.000 & 0.220& &0.000& 0.000&0.060&\\
&  &38&16&-0.050 & 0.000 & 0.160& &0.001& 0.000&0.010&\\ 
&  &40&18& 0.000 & 0.000 & 0.140& &0.000& 0.000&0.020&\\
&  &42&20& 0.000 & 0.000 & 0.000& &0.000& 0.000&-0.010&\\
&  &44&22& 0.000 & 0.000 & 0.160& &0.000& 0.000&0.050&\\
&  &46&24& 0.112 & 0.000 & 0.220& &0.027& 0.000&0.060&\\ 
&  &48&26& 0.000 & 0.000 & 0.170& &0.000& 0.000&0.020&\\  
&  &50&28& 0.000 & 0.000 & 0.080& &0.000& 0.000&0.010&\\
&  &52&30& 0.000 & 0.000 & 0.140& &0.000& 0.000&0.000&\\ 
&  &54&32& 0.120 & 0.000 & 0.020& &0.012& 0.000&0.000&\\
&  &56&34& 0.122 & 0.135 & 0.020& &-0.006&-0.018&0.000&\\
&  &58&36&-0.089 &-0.105 & 0.070& &-0.005&-0.011&-0.010&\\ 
&  &60&38&-0.001 &-0.079 & 0.060& & 0.000&-0.029&0.000&\\
&  &62&40& 0.000 & 0.000 & 0.020& & 0.000&0.000&0.000&\\
&  &64&42& 0.000 & 0.027 & 0.230& & 0.000&0.000&0.050&\\
&  &66&44& 0.001 & 0.147 & 0.200& & 0.000&0.108&0.050&\\
&  &68&46& 0.146 & 0.152 & 0.190& & 0.048&0.067&0.040&\\
&  &70&48& 0.112 & 0.099 & 0.170& & 0.020&-0.021&0.000&\\
&  &72&50& 0.000 & 0.045 &-0.090& & 0.000&0.001&0.030&\\
&  &74&52& 0.007 & 0.119 &-0.130& & 0.000&0.105&0.030&\\    
\hline
&24&38&14& 0.287 & 0.234 &-     & & 0.061  & 0.139&-    &\\
&  &40&16& 0.262 & 0.273 &0.300 & & 0.014  & 0.027&0.040 &\\  
&  &42&18& 0.124 & 0.000 &0.240 & & 0.003  & 0.000&0.030 &\\
&  &44&20& 0.000 & 0.000 &0.150 & & 0.000  & 0.000&0.030 &\\ 
&  &46&22& 0.114 & 0.000 &0.240 & & 0.030  & 0.000&0.060&\\ 
&  &48&24& 0.248 & 0.000 &0.260 & & 0.072  & 0.000&0.070&\\  
&  &50&26& 0.204 & 0.000 & 0.240& & 0.025  & 0.000& 0.030&\\
&  &52&28& 0.000 & 0.000 & 0.140& & 0.000  & 0.000& 0.020&\\
&  &54&30& 0.177 & 0.180 & 0.180& & 0.034  & 0.045& 0.000&\\ 
&  &56&32& 0.212 & 0.189 & 0.130& & 0.015  & 0.022& 0.010&\\ 
&  &58&34& 0.204 & 0.199 & 0.150& &-0.020  &-0.026& 0.000&\\
&  &60&36& 0.153 & 0.181 & 0.170& &-0.021  &-0.021& 0.020&\\
&  &62&38&-0.047 & 0.329 & 0.280& &-0.005  & 0.047& 0.060&\\
&  &64&40& 0.000 & 0.018 & 0.280& & 0.000  & 0.000& 0.060&\\
&  &66&42& 0.000 & 0.053 & 0.270& & 0.000  & 0.009& 0.050&\\ 
&  &68&44& 0.179 & 0.161 & 0.260& & 0.069  & 0.084& 0.040&\\ 
&  &70&46& 0.200 & 0.169 & 0.210& & 0.054  & 0.062& 0.040&\\
&  &72&48& 0.188 & 0.126 & 0.180& & 0.028   &-0.019&0.010&\\ 
&  &74&50& 0.002 & 0.053 & 0.020& & 0.000   & 0.001&0.000&\\
&  &76&52& 0.036 & 0.142 & 0.140& & 0.003   & 0.091&0.020&\\ 
&  &78&54& 0.149 & 0.178 & 0.130& & 0.020   & 0.072&-0.020&\\
&  &80&56& 0.162 & 0.254 &  -   & & 0.003   & 0.152& - &\\ 
\hline
\end{tabular}
\end{center}
\end{table}
\vfill
\noindent\begin{table}
\begin{center}
\caption{\sf The quadrupole deformations $\beta_2$ for various nuclei
in the RMF theory. The available empirical deformations (expt.) 
obtained from BE(2) values {\protect \cite{Ram.87}} are also given. The 
experimental values for a few S (A=38-42) and Ar (A=44,46) 
isotopes are taken from the recent measurements {\protect \cite{Scheit.96}}. 
The experimental values do not show the sign of the deformation.}
\bigskip
\begin{tabular}{lll  c c c c c c c  l}
\hline\hline 
& Nucleus  & RMF & expt.& & Nucleus & RMF & expt. & \\
\hline
& $^{18}$Ne&0.002&  0.691 &   &  $^{42}$S &0.259 & 0.300 & \\
& $^{20}$Ne&0.241& 0.728  &   &  $^{34}$Ar  & 0.162&0.238    &\\
& $^{22}$Ne&0.408& 0.562  &   &  $^{36}$Ar  &-0.195&0.273     &\\
& $^{24}$Ne&0.178& 0.410  &   &  $^{38}$Ar & 0.000& 0.162   &  \\
& $^{22}$Mg&0.432& 0.560  &   &  $^{40}$Ar  & 0.000& 0.251 &\\
& $^{24}$Mg&0.465& 0.606  &   &  $^{42}$Ar  & 0.122& 0.273 &  \\
& $^{26}$Mg&0.298& 0.482  &   &  $^{44}$Ar  & 0.122& 0.241 &  \\
& $^{28}$Mg&0.288& 0.485  &   &  $^{46}$Ar  & 0.000& 0.176 &  \\
& $^{26}$Si&0.329  & 0.444  &  & $^{42}$Ti  & 0.000& 0.310     &\\ 
& $^{28}$Si&-0.287 & 0.407  &  & $^{44}$Ti  & 0.000& 0.262     &\\
& $^{30}$Si&-0.185 & 0.316  &  & $^{46}$Ti& 0.112& 0.317     &  \\
& $^{32}$Si&0.028  & 0.345  &  & $^{48}$Ti  & 0.000 &0.269    & \\
& $^{30}$S &-0.200 & 0.336  & & $^{50}$Ti  & 0.000& 0.166    &\\
& $^{32}$S &0.235& 0.312  &   &  $^{48}$Cr & 0.248& 0.335    &  \\
& $^{34}$S &0.146& 0.252  &   &  $^{50}$Cr  & 0.204&0.293    & \\
& $^{36}$S &0.000& 0.164  &   & $^{52}$Cr  & 0.000 &0.224   &\\
& $^{38}$S &-0.043 & 0.246 &   &  $^{54}$Cr  & 0.177&0.250    &\\
& $^{40}$S &0.242 & 0.284 &   &              &      &         & \\    
\hline\hline
\end{tabular}
\end{center}
\end{table}
\newpage

\noindent\begin{table}
\begin{center}
\caption{\sf Nuclei with a shape coexistence in the ground state.
The difference in the energy of the ground state in the prolate
and oblate shape is given. The associated deformations are also
shown.}
\bigskip
\begin{tabular}{ll c c c c c l}
\hline\hline
&Nucleus & $E_{pro}-E_{obl}$ & $\beta_2 (pro.)$ & $\beta_2 (obl.)$ & \\
\hline
&$^{24}$Ne & $-$0.170 & 0.177  &$-$0.107 &\\
&$^{26}$Mg & $-$0.370 & 0.300  &$-$0.219 &\\
&$^{30}$Mg & $-$0.406 & 0.190  &$-$0.133 &\\
&$^{42}$Mg & $-$1.374 & 0.410  &$-$0.300 &\\
&$^{44}$Mg & $-$0.236 & 0.320  &$-$0.240 &\\
&$^{26}$Si & $-$0.232 & 0.330  &$-$0.230 &\\
&$^{30}$Si & $-$0.110 & 0.120  &$-$0.185 &\\
&$^{32}$Si &    0.455 & 0.030  &$-$0.150 &\\
&$^{38}$Si & $-$0.790 & 0.250  &$-$0.155 &\\
&$^{40}$Si & $-$0.170 & 0.290  &$-$0.240 &\\
&$^{44}$S  &    0.030 & 0.375  &$-$0.190 &\\
&$^{48}$S  & $-$0.160 & 0.200  &$-$0.200 &\\
&$^{50}$S  &    0.208 & 0.170  &$-$0.220 &\\
&$^{52}$S  &    0.145 & 0.130  &$-$0.170 &\\
&$^{54}$S  &    0.026 & 0.078  &$-$0.087 &\\
&$^{30}$Ar & $-$0.650 & 0.230  &$-$0.140 &\\
\hline\hline
\end{tabular}
\end{center}
\end{table}
\newpage

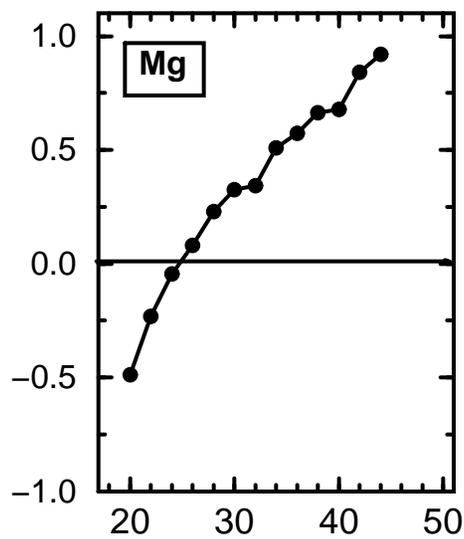
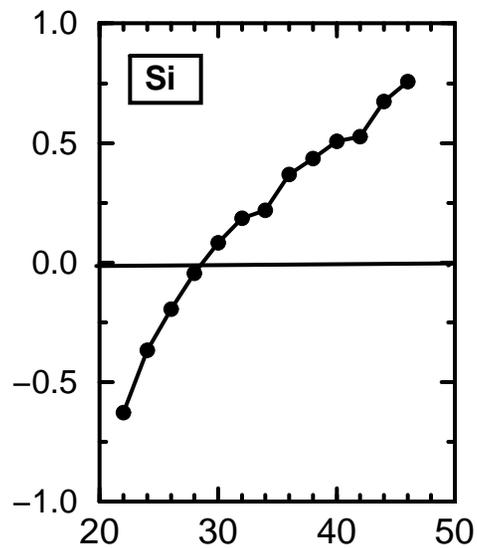
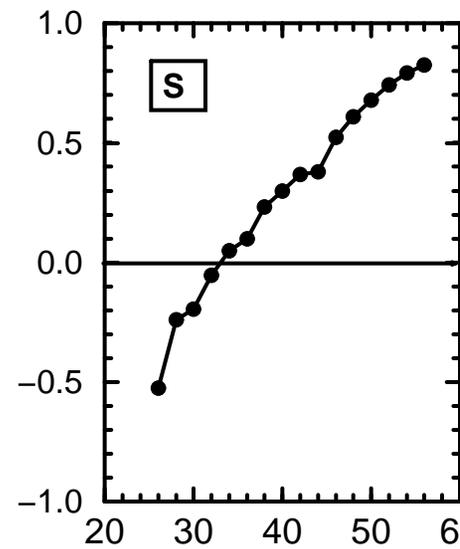
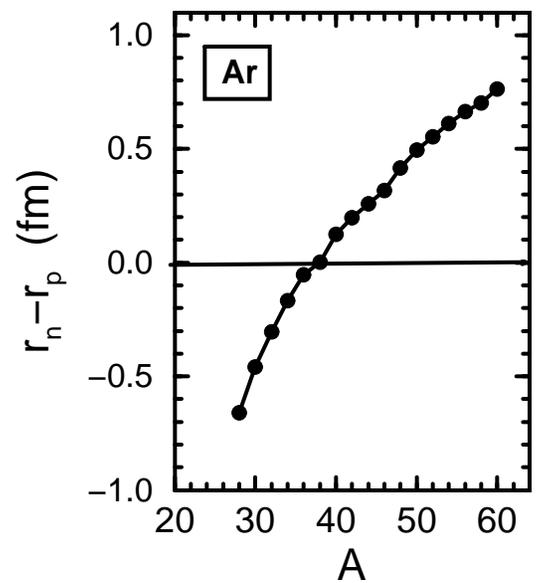
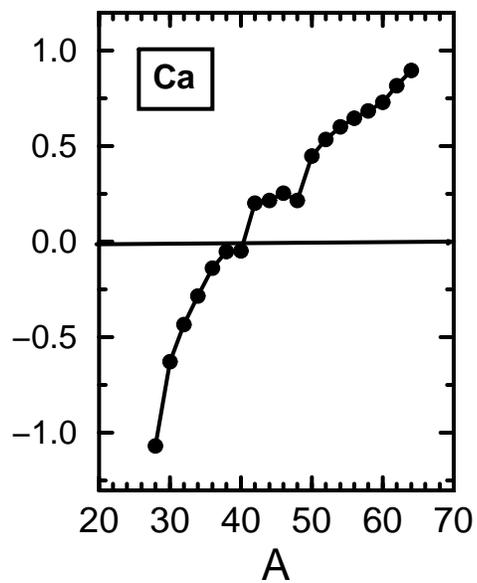
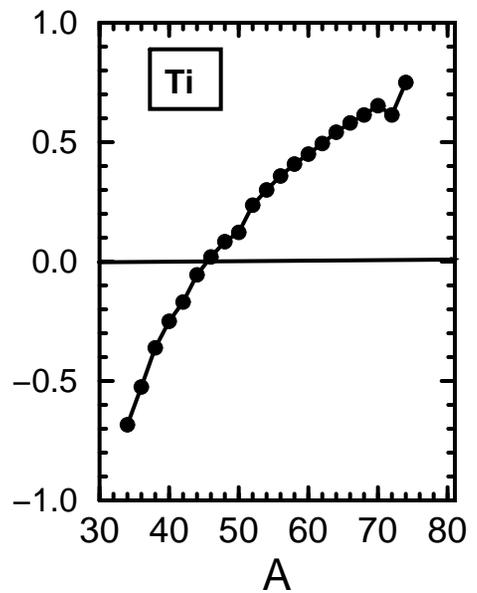
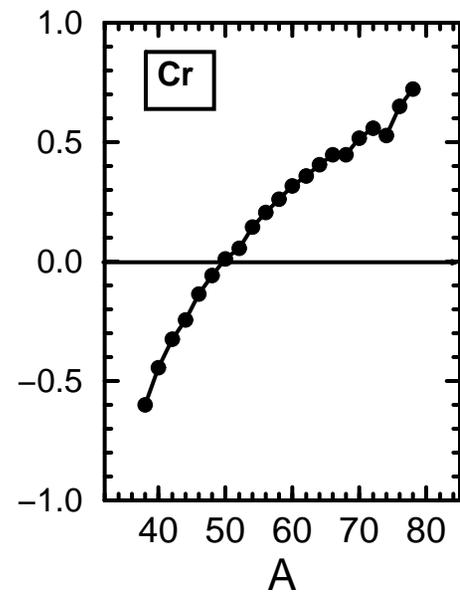
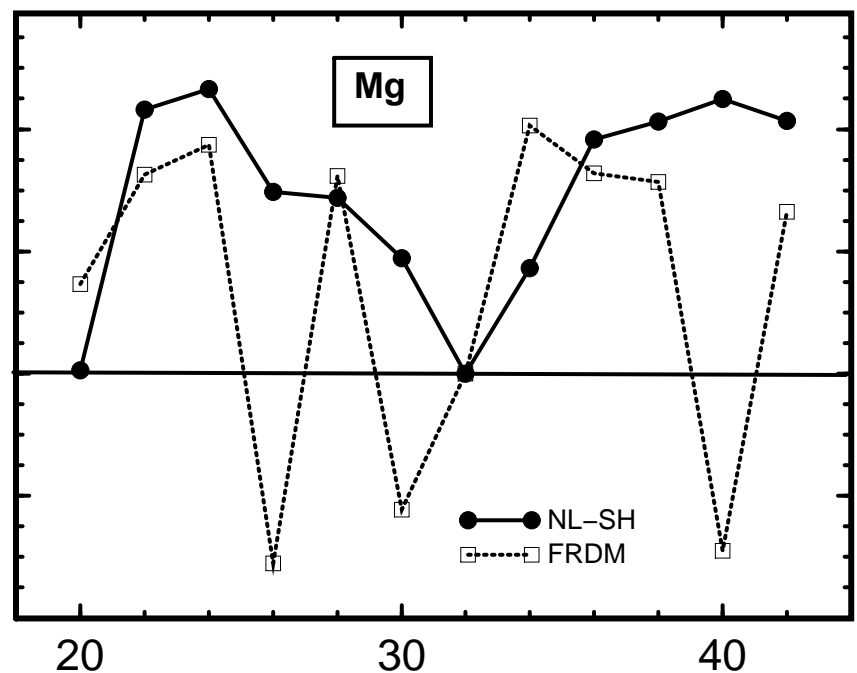
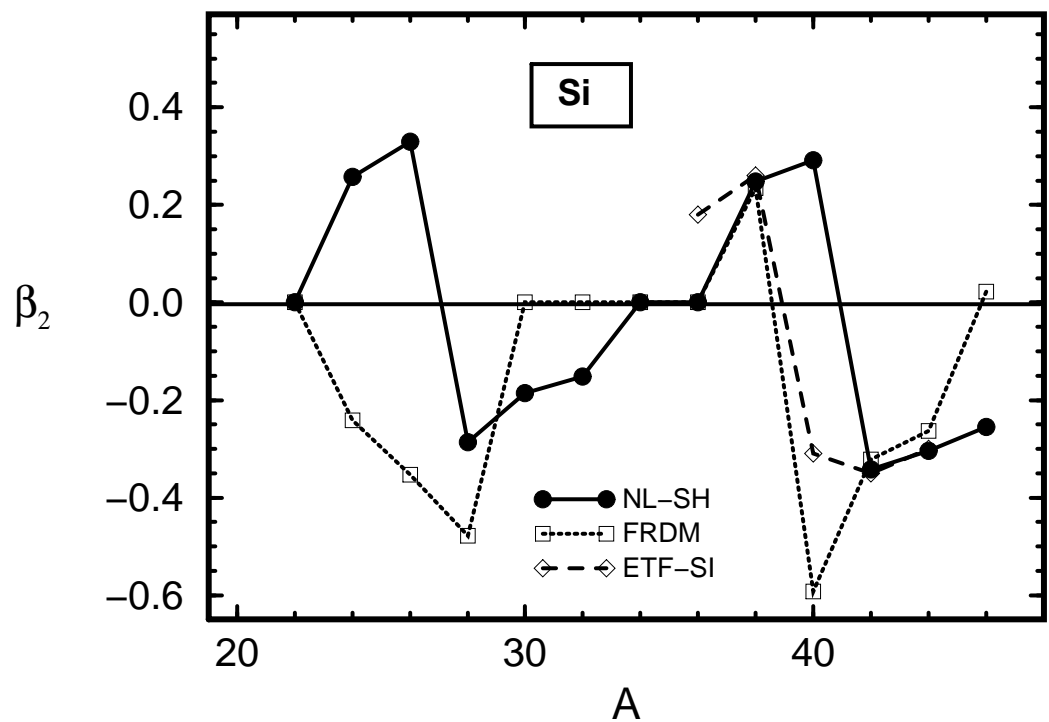
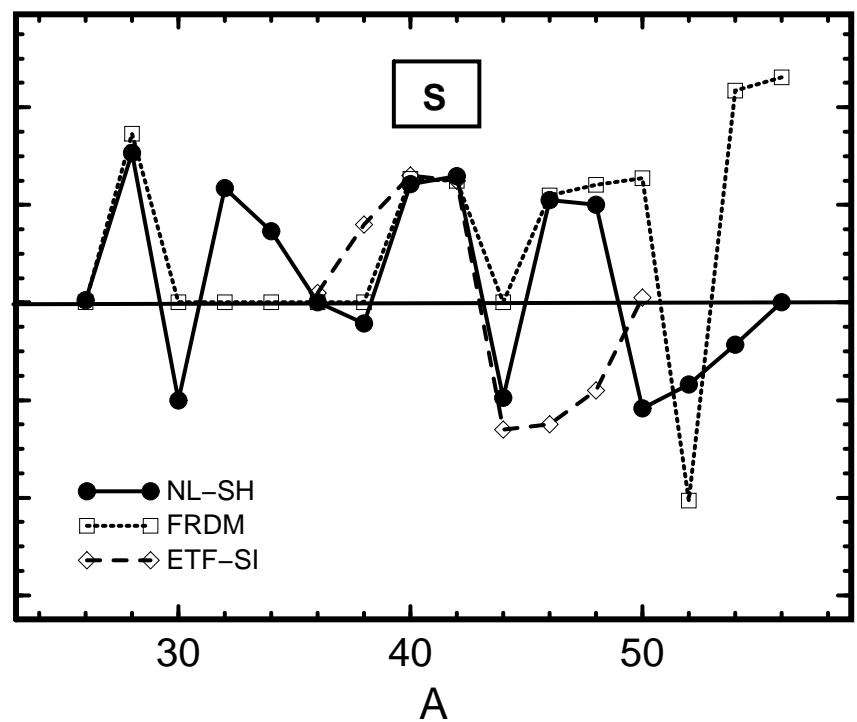
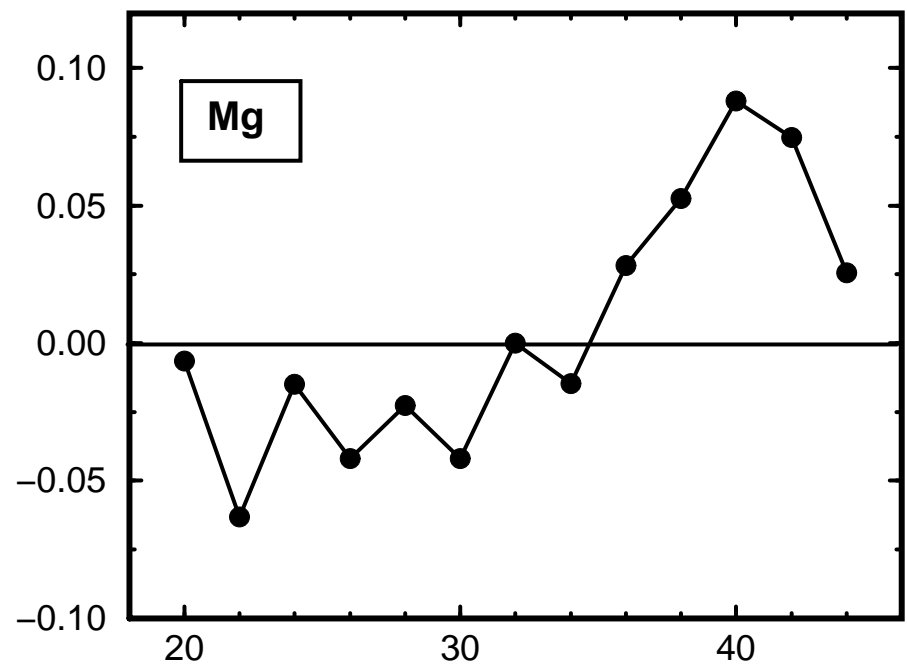
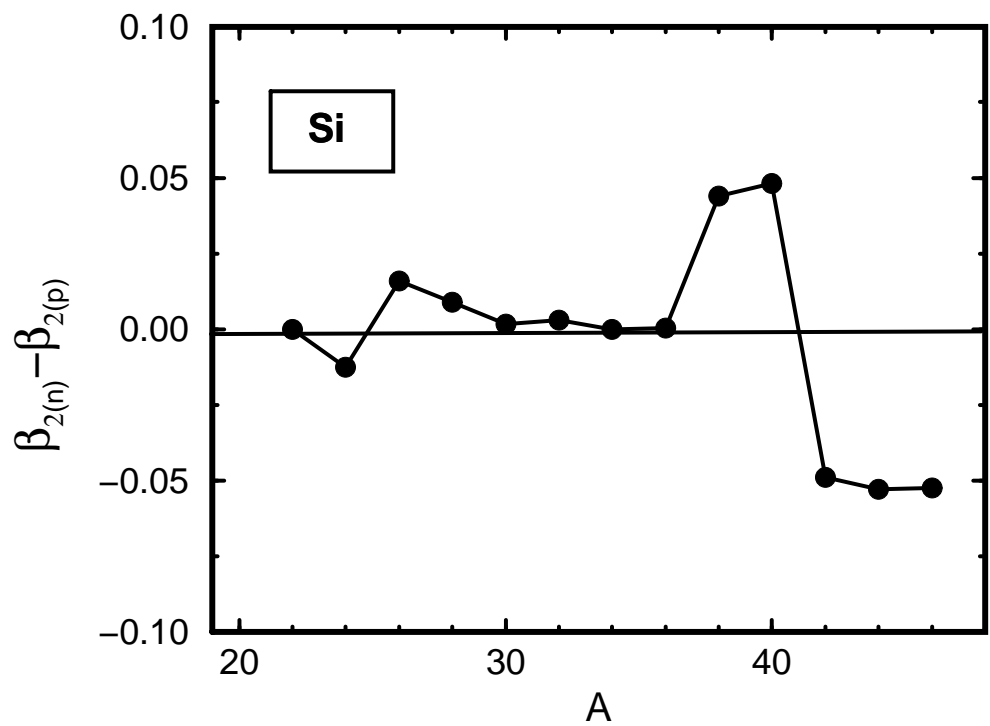
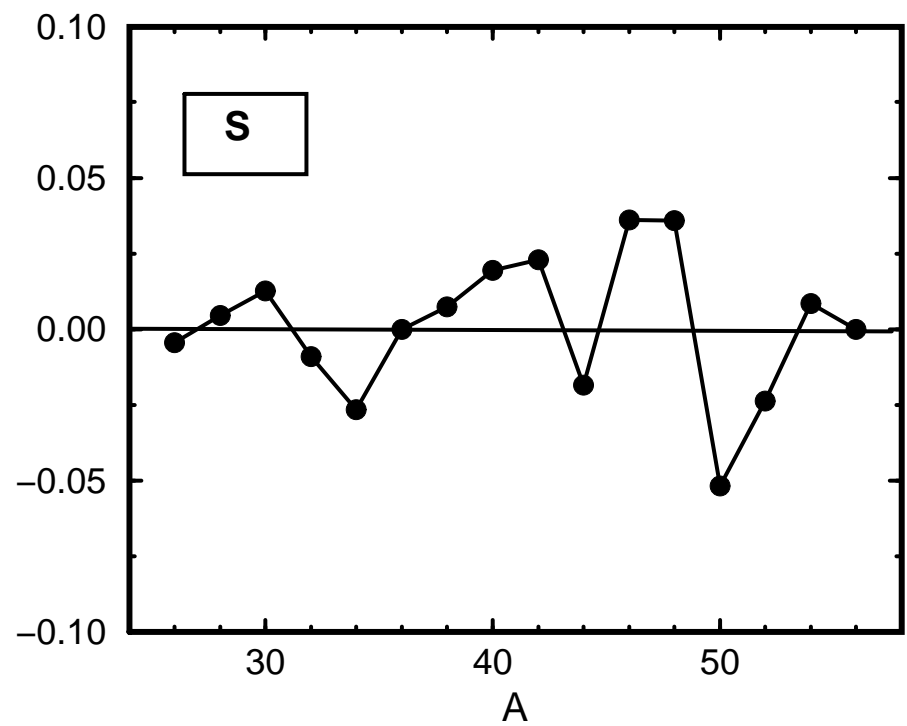
\end{document}